\documentclass[aps,pre,superscriptaddress,citeautoscript, amsmath,amssymb,twocolumn]{revtex4-1}	
\usepackage{graphicx}
\usepackage{amsmath}
\usepackage{hyperref}
\usepackage{color}
\usepackage{bigints}
\usepackage{mathrsfs}
\usepackage{algorithm}
\usepackage{algpseudocode}
\usepackage{multirow}
\usepackage[utf8]{inputenc}

\definecolor{darkblue}{RGB}{8,81,156}
\hypersetup{colorlinks,breaklinks,
            linkcolor=darkblue,urlcolor=darkblue,
           anchorcolor=darkblue,citecolor=darkblue}

\date{\today}

\allowdisplaybreaks

    \definecolor{dark-purple}{RGB}{118,42,131}
    \definecolor{dark-green}{RGB}{27,120,55}
    \definecolor{light-purple}{RGB}{231,212,232}
    \definecolor{LIGHT-PURPLE}{RGB}{194,165,207}
    \definecolor{light-green}{RGB}{168,216,183}
    \definecolor{gray}{RGB}{186,186,186}
    \definecolor{super-dark-green}{RGB}{0,69,41}
    \definecolor{super-dark-purple}{RGB}{63,0,125}
    \definecolor{super-dark-blue}{RGB}{8,48,107}
    \definecolor{super-dark-red}{RGB}{165,0,38}
    
    \definecolor{super-dark-purple}{RGB}{64,0,75}
    \definecolor{super-dark-green}{RGB}{0,68,27}

\usepackage{tabularx,booktabs} 

\usepackage{array}
\usepackage{longtable}
\newcolumntype{L}[1]{>{\raggedright\let\newline\\\arraybackslash\hspace{0pt}}p{#1}}
\newcolumntype{C}[1]{>{\centering\let\newline\\\arraybackslash\hspace{0pt}}m{#1}}
\newcolumntype{R}[1]{>{\raggedleft\let\newline\\\arraybackslash\hspace{0pt}}m{#1}}

\setcitestyle{super}

\usepackage{atveryend}
\makeatletter
\let\origcitation\citation
\AtEndDocument{\def\mycites{\@gobble}%
  \def\citation#1{\g@addto@macro\mycites{,#1}\origcitation{#1}}}
\AtVeryEndDocument{\typeout{***^^JCited keys: \mycites^^J***}}
\makeatother

\makeatletter
\let\origcitation\citation
\AtEndDocument{\def\mycites{}%
  \def\citation#1{\g@addto@macro\mycites{#1^^J,}\origcitation{#1}}}
\AtVeryEndDocument{\newwrite\citeout\immediate\openout\citeout=\jobname.cit
  \immediate\write\citeout{\mycites}\immediate\closeout\citeout}
\makeatother

\begin{document}

\title{Role of Nanoscale Interfacial Proximity in Contact Freezing in Water}

\author{Sarwar Hussain}
\email{sarwar.hussain@yale.edu}
\affiliation{Department of Chemical and Environmental Engineering, Yale University, New Haven, CT  06520}

\author{Amir Haji-Akbari}
\email{amir.hajiakbaribalou@yale.edu}
\affiliation{Department of Chemical and Environmental Engineering, Yale University, New Haven, CT  06520}

\begin{abstract}
Contact freezing is a mode of atmospheric ice nucleation  in which a collision between a dry ice nucleating particle (INP) and a water droplet results in considerably faster heterogeneous  nucleation. The molecular mechanism of such enhancement is, however, still a mystery. While earlier studies had attributed it to collision-induced transient perturbations, recent experiments point to the pivotal role of nanoscale proximity of the INP and the free interface. By simulating   heterogeneous  nucleation of ice within INP-supported nanofilms of two model water-like tetrahedral liquids, we demonstrate that such nanoscale proximity is  sufficient for inducing rate increases commensurate with those observed in contact freezing experiments, but only  if the free interface has a tendency to enhance homogeneous nucleation. Water is suspected of possessing this latter property, known as surface freezing propensity. Our findings therefore establish a connection between surface freezing propensity and kinetic enhancement during contact nucleation. We also observe that faster nucleation proceeds through a mechanism markedly distinct from classical heterogeneous nucleation, involving the formation of hourglass-shaped crystalline nuclei that conceive at either interface, and that have a lower free energy of formation due to the nanoscale proximity of the  interfaces and the modulation of the free interfacial structure by the INP. In addition to providing valuable insights into the physics of contact nucleation, our findings can assist in improving the accuracy of heterogeneous nucleation rate measurements in experiments, and in advancing our understanding of ice nucleation on nonuniform  surfaces such as organic, polymeric and biological materials. 
\end{abstract}

\maketitle

\section{Introduction}

\noindent
Crystallization of liquids under nanoscale confinement has received considerable attention in recent decades~\cite{GiovambattistaAnnRevPhysChem2012}, as drastic changes in the thermodynamics of freezing~\cite{TakaiwaPNAS2008, HanNaturePhys2010, AgrawalNatureNano2016, GaoScientificReports2018, KumarLangmuir2018, JantschJPCC2019}, the kinetics and mechanism of nucleation~\cite{JungwithJPCB2006, GalliNatComm2013, HajiAkbariFilmMolinero2014, GianettiPCCP2016, HajiAkbariPNAS2017,  CampbellPNAS2017, MascottoJPCC2017, BiNatComm2017, CampbellPRL2018}, and the identity of the nucleated crystals~\cite{KogaNature2001, TakaiwaPNAS2008, GaoScientificReports2018} have been reported in numerous experimental~\cite{AgrawalNatureNano2016, CampbellPNAS2017, MascottoJPCC2017, CampbellPRL2018, JantschJPCC2019} and computational~\cite{KogaNature2001, JungwithJPCB2006, TakaiwaPNAS2008, HanNaturePhys2010, GalliNatComm2013, HajiAkbariFilmMolinero2014, GianettiPCCP2016,  HajiAkbariPNAS2017, BiNatComm2017, GaoScientificReports2018} studies of freezing in nanopores~\cite{CampbellPNAS2017, MascottoJPCC2017, CampbellPRL2018, JantschJPCC2019}, nanotubes~\cite{KogaNature2001, TakaiwaPNAS2008,  AgrawalNatureNano2016}, slit pores~\cite{HanNaturePhys2010, KumarLangmuir2018}, wedges~\cite{BiNatComm2017}, nanodroplets~\cite{GalliNatComm2013} and freestanding nanofilms~\cite{JungwithJPCB2006, HajiAkbariFilmMolinero2014, GianettiPCCP2016,  HajiAkbariPNAS2017}. What has received less attention is the freezing of liquids under mixed-interface confinement, i.e.,~when a liquid is sandwiched between a solid-liquid and a free interface. Mixed-interface confinement can emerge in many different environments, and can dramatically impact the spatial heterogeneity and kinetic stability of the corresponding systems.  Examples include the glass transition temperatures\cite{FrankScience1996, EdigerScience2007} and crystallization tendencies\cite{GiriNatComm2014} of organic and polymeric films, and the formation of low-dimensional ices on solid surfaces\cite{OdeliusPRL1997, KimmelJACS2009, XuScience2010, ZhengAngewChem2013}. One
  process that can be strongly affected by mixed-interface confinement is atmospheric ice nucleation, which plays a pivotal role in cloud microphysics~\cite{MurrayGRL2015}. The dominant mode of ice formation in clouds is \emph{immersion freezing} (Figure~\ref{Figure 1}a) in which ice nucleates heterogeneously on an ice nucleating particle (INP) fully immersed within an atmospheric microdroplet~\cite{KnopfACSEarthSpaceChem2018}. INPs can, however,  come into close proximity of free interfaces, which can, in turn, alter the kinetics and mechanism of nucleation in nontrivial ways. Considering the highly stochastic nature of immersion freezing~\cite{KnopfClimAtmosSci2020}, such changes can, for instance, introduce large uncertainties into experimental estimates of immersion nucleation rates. 

A more intriguing example is \emph{contact freezing} {(Figure~\ref{Figure 1}b-c) in which nucleation is triggered by a collision between a dry INP and a supercooled water droplet~\cite{RauZeitNaturforsch1950}. Contact freezing is usually orders of magnitude faster than immersion freezing as demonstrated for a wide variety of INPs~\cite{DurantGeophysResLett2005, ShawJPCB2005, ForneaJGeophysRes2009, LadinoJGeophysRes2011,  HoffmannAtmosMeasTechDiscuss2013}. Pinpointing the molecular origin of this enhancement is, however, extremely difficult due to the transient nature of contact freezing.  While earlier works had mostly attributed it to transient factors caused by the collision\cite{FletcherJAtmosSci1970, CooperJAtmosSci1974, FukutaJAtmosSci1975, NiehausJPhysChemLett2015}, more recent studies point to the pivotal role of mixed-interface confinement.  For instance, Shaw and Durant~\cite{DurantGeophysResLett2005, ShawJPCB2005} observed that the kinetic freezing temperature in droplets that undergo repeated cycles of freezing and melting only depends on the proximity of the INP and the free interface, and is independent of whether the INP approaches the interface from outside (Figure~\ref{Figure 1}c) or from within (Figure~\ref{Figure 1}b). These two modes of contact freezing are referred to as ''outside-in`` and ''inside-out`` freezing, respectively.  Initially, it was argued that these observations might be explained by the supposed tendency of the water-vapor-INP contact line to facilitate nucleation~\cite{SearJPhysCondensMatter2007, DjikaevJPhysChemA2008p}.  Later experiments by the same group, however, found no such tendency~\cite{GurganusJPhysChemLett2011, GurganusJPhysChemC2013} except for INPs with nanoscale texture~\cite{GurganusPRL2014}, suggesting that nanoscale proximity might result in faster nucleation even in the absence of a contact line.

\begin{figure}[h]
\centering
\includegraphics[width=0.4365\textwidth]{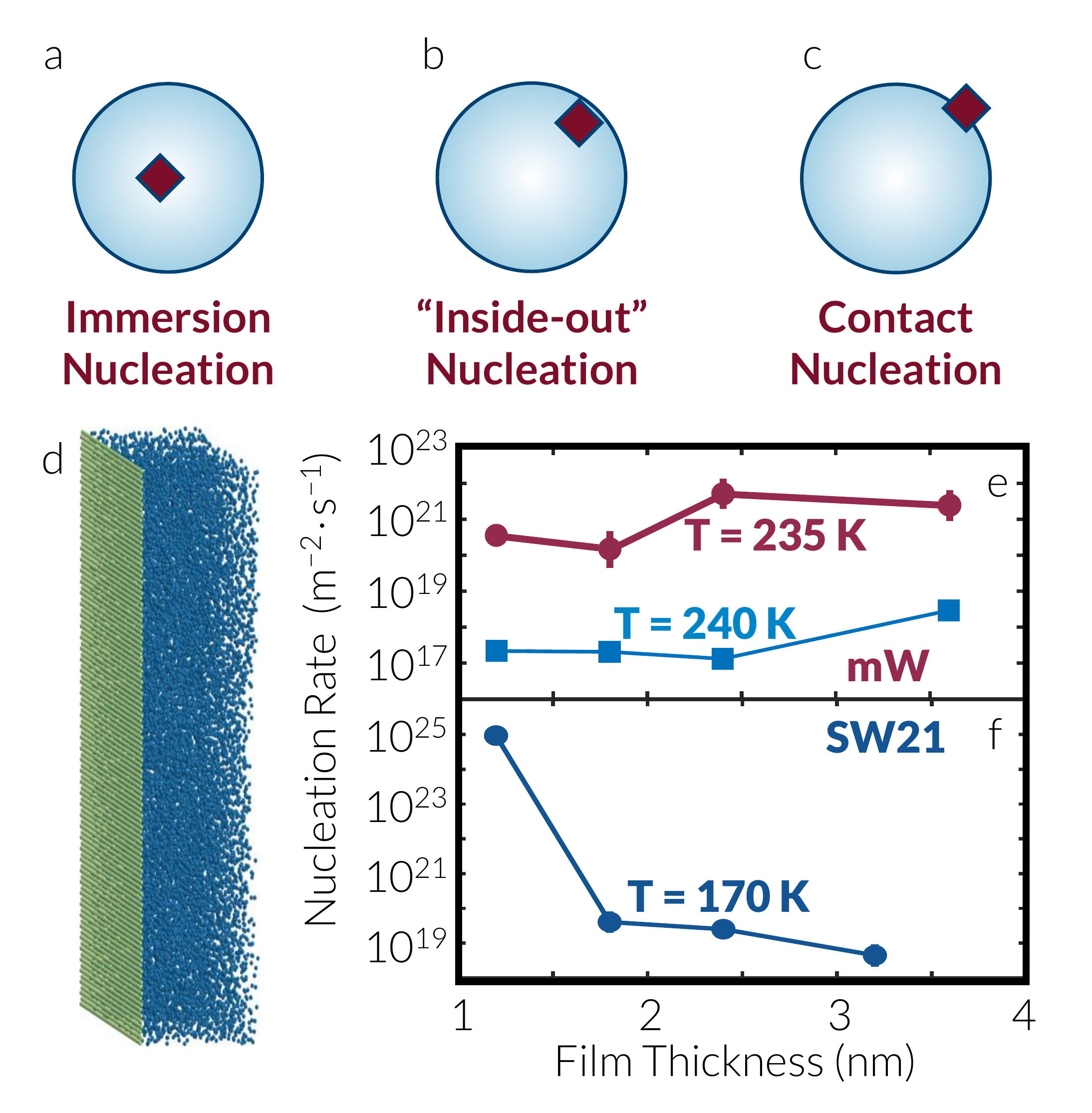} 
\caption{\textbf{Free Interfaces and the Kinetics of Heterogeneous Ice Nucleation} (a-c) A schematic representation of (a) immersion freezing, and (b) inside-out and (c) conventional contact freezing, the three modes of heterogeneous ice nucleation discussed in this work. The INPs and water droplets are depicted in dark red and light blue, respectively. (d) A schematic representation of a graphene-supported thin film, with water molecules and carbon atoms depicted in dark blue and light green, respectively. (e-f) The dependence of heterogenous nucleation rate on film thickness in (e) mW and (f) SW21 films. Error bars correspond to 95\% confidence intervals and are smaller than the symbols.}
\label{Figure 1}
\end{figure}

Despite these remarkable findings, there is a considerable gap in our understanding of how a vapor-liquid interface can enhance heterogeneous nucleation on a proximal INP. It has been argued~\cite{DurantGeophysResLett2005, ShawJPCB2005, DjikaevJPhysChemA2008p} that this tendency might be linked to the suspected ability of a free interface to facilitate homogeneous ice nucleation. Here, ''homogeneous nucleation`` refers to nucleation in the absence of an extrinsic INP, whether it occurs in the bulk or is facilitated at the vapor-liquid interface. Surface-dominated homogeneous nucleation-- typically referred to as \emph{surface freezing}-- was originally proposed by Tabazadeh~\emph{et al.}~\cite{TabazadehPNAS2002} and has since been extensively studied experimentally~\cite{DuftACPD2004, CiobanuJPhysChemA2010, KuhnAtmosChemPhys2011} and computationally~\cite{JungwithJPCB2006, GalliNatComm2013, HajiAkbariFilmMolinero2014, GianettiPCCP2016, HajiAkbariPNAS2017}. Experimental evidence for surface freezing, however, is inconclusive mostly due to the difficulty of generating mono-dispersed droplets in the sub-$\mu$m size regime where this effect is predicted to become dominant~\cite{SignorellPhysRevE2008}. Computational studies of surface freezing have also been equally inconclusive~\cite{HajiAkbariJCP2017}, as the enhancement of nucleation is only observed~\cite{JungwithJPCB2006, GianettiPCCP2016,  HajiAkbariPNAS2017} for some force-fields, such as the atomistic TIP4P/Ice~\cite{VegaTIP4PiceJCP2005} model, and not others~\cite{GalliNatComm2013, HajiAkbariFilmMolinero2014, GianettiPCCP2016}, such as the coarse-grained monoatomic water (mW)~\cite{MolineroJPCB2009} potential. This force-field dependence of surface freezing propensity provides an opening for testing the hypothesized connection between free surface-induced homogeneous and heterogeneous nucleation, or surface and contact freezing, respectively. If such a relationship exists, the rate of heterogeneous nucleation in supported liquid nanofilms of a model prone to surface freezing will decrease drastically with film thickness, while no (or the opposite) dependence on thickness will be observed for the force-field(s) with no surface freezing propensity. 

Here, we use molecular dynamics simulations and our recently developed jumpy forward flux sampling (jFFS) algorithm~\cite{HajiAkbariJCP2018} to test this hypothesis by computing the rates and characterizing the mechanism of heterogeneous nucleation  in supported nanofilms of two model water-like tetrahedral liquids with opposing surface freezing propensities. Our focus on supported nanofilms not only allows us to test this hypothesized connection, but also enables us to probe the exclusive effect of interface proximity on heterogeneous nucleation, in the absence of other competing factors such as contact lines, interfacial curvature and collision-induced perturbations.   In other words, supported films constitute ideal model systems for determining whether nanoscale proximity is sufficient for inducing faster heterogeneous nucleation in contact freezing. As such, we will refer to strong sensitivity of rate to film thickness as ''contact freezing propensity`` for brevity, even though contact freezing is a complex phenomenon whose precise kinetics and mechanism is likely impacted by a plethora of other factors. Our calculations reveal that nanoscale proximity is indeed a sufficient condition for inducing kinetic enhancements of the types observed in contact freezing experiments, but only for the liquid that is prone to surface freezing.  More precisely, heterogeneous nucleation  becomes orders of magnitude faster in ultrathin films of the surface-freezing liquid wherein critical nuclei adopt a hourglass-shaped structure due to nanoscale proximity of the two  interfaces, while no dependence  of rate on thickness is observed for the other liquid.  Our analysis using classical nucleation theory (CNT)~\cite{TurnballJCP1950} reveals that the formation of such nuclei can result in a decrease in nucleation barrier, but not by enough to quantitatively explain the observed increase in rate. We explain this discrepancy by noting that the presence of an INP modulates the free interfacial structure of the films exhibiting faster nucleation, which results in a decrease in the effective contact angle and the nucleation barrier.

\section{Methods}\label{Computational Method}

\noindent\textbf{System Description and Molecular Dynamics Simulations:} 
We consider supported films of two water-like tetrahedral liquids. Both liquids belong to the Stillinger-Weber (SW)~\cite{StillingerPRB1985} family of potentials in which the tetrahedral arrangement of nearest neighbors around a central site is enforced by including in the interatomic potential  a three-body term that penalizes deviations from the tetrahedral angle. The magnitude of the energetic penalty is tuned using a parameter called tetrahedrality, $\lambda$. The first model liquid is mW~\cite{MolineroJPCB2009}, a widely used coarse-grained model of water with  $\lambda=23.15$, which has been shown~\cite{GalliNatComm2013, HajiAkbariFilmMolinero2014} to not undergo surface freezing, while the second liquid is a re-parameterized variant of mW with  $\lambda=21$ that undergoes~\cite{GianettiPCCP2016} surface freezing.
We call this second liquid SW21, which is different from real water and mW in several aspects such as its melting point (206~K for SW21 vs.~274~K for mW) and its hydration structure~\cite{GianettiPCCP2016}. The precise phase diagrams of these two models can be found elsewhere~\cite{MolineroJPCB2009, DhabalJChemPhys2015}. We choose SW21 over atomistic models with surface freezing propensity-- such as TIP4P/Ice~\cite{VegaTIP4PiceJCP2005}-- not only due to the prohibitively large computational cost of the latter, but also because comparing SW21 and mW allows us to explore the effect of surface freezing propensity on heterogeneous nucleation in two models that are otherwise similar.
Conversely, comparing any potential differences between the contact freezing propensities of TIP4P/Ice and mW could not be conclusively attributed to their differing surface freezing tendencies, and could instead be caused by other factors, such as the presence of electrostatic interactions in the TIP4P/Ice system. The temperatures at which rates are computed  correspond to similar relative supercoolings (or $T/T_m$ values), which all lie between 0.8 and 0.87. We put liquid films of mW and SW21 (Figure~\ref{Figure 1}d) in contact with two types of model INPs. The first INP is a graphene wall that interacts with liquid molecules via the two-body part of the SW potential, with $\epsilon_{\text{mW}}^g=0.52~\text{kcal}\cdot\text{mol}^{-1}$ and $\epsilon_{\text{SW21}}^g=0.13~\text{kcal}\cdot\text{mol}^{-1}$ for mW and SW21, respectively.  $\epsilon_{\text{mW}}^g$ is adopted from Bi~\emph{et al.}~\cite{BiJPhysChemC2016}, while $\epsilon_{\text{SW21}}^g$ is chosen because no heterogeneous nucleation is observed for $\epsilon_{\text{mW}}^g$ in the SW21 system. We use a value of $\sigma^g=0.32$~nm for both liquids.  The second INP is a structureless attractive wall interacting via the Lennard-Jones (LJ) 9-3 potential~\cite{MagdaJCP1985} with $\epsilon_{\text{mW}}^{\text{LJ}}=1.2~\text{kcal}\cdot\text{mol}^{-1}$ and $\sigma_{\text{mW}}^{\text{LJ}}=0.32$~nm,  and $\epsilon_{\text{SW21}}^{\text{LJ}}=0.48~\text{kcal}\cdot\text{mol}^{-1}$ and $\sigma_{\text{SW21}}^{\text{LJ}}=0.3$~nm for mW and SW21, respectively. These values represent the smallest $\epsilon$'s for which heterogeneous nucleation is observed in conventional 50-ns long MD simulations at 215~K and 155~K in the mW and SW21 systems, respectively. All LJ 9-3 interactions are truncated at $2.5\sigma$. 

All MD simulations are performed in the canonical (NVT) ensemble using the large-scale atomic/molecular massively-parallel simulator (\textsc{LAMMPS})~\cite{PimptonLAMMPS1995} package. Equations of motion are integrated  using the velocity-Verlet algorithm with a time step of 5~fs, while temperature is controlled using the Nos\'{e}-Hoover~\cite{NoseMolPhys1984, HooverPhysRevA1985} thermostat with a time constant of 0.5~ps. Supercooled liquid configurations are prepared by melting a properly-sized film of cubic ice at 350~K and 250~K for mW and SW21 systems, respectively. We collect a minimum of 100 melted configurations once every $0.05$~ns, and gradually quench them to the respective target temperature at  a cooling rate of $6.25~\text{ps}\cdot\text{K}^{-1}$. This choice is guided by the fact that structural relaxation times for both mW and SW21 never exceed 2~ps within the range of temperatures considered in this work (Figure~\ref{fig:relaxation-times}). Therefore, the quenched configurations have sufficient time to structurally relax during cooling and do not get kinetically arrested. Note that the ensuing films are sandwiched between the INP and the vapor phase, which, due to the low vapor pressure of mW-like models under supercooled conditions~\cite{FactorovichJCP2014}, is technically indistinguishable from vacuum. This implies simulating nucleation at zero pressure, which accurately represents  atmospherically relevant conditions. All system characteristics (including system sizes) are given in Table~\ref{tab:system-summary}.

\noindent\textbf{Rate Calculations:}
Nucleation rates are computed using our recently developed jFFS algorithm~\cite{HajiAkbariJCP2018} with the number of molecules within the largest crystalline nucleus as the order parameter, $\xi(\cdot)$. Individual molecules are classified as solid-like or liquid-like based on the $q_6$ Steinhart bond order parameter~\cite{SteinhardtPRB1983}, and the solid-like molecules within a distance cutoff $r_c$ are clustered to form crystalline nuclei. In order to be consistent with our earlier work on the SW21 model~\cite{GianettiPCCP2016}, we use $r_c=0.32$~nm and $0.345$~nm for mW and SW21, respectively, and apply the chain exclusion algorithm of Reinhardt~\emph{et al.}~\cite{VegaJCP2012}. Further details about the particular definition of $q_6$ and the clustering algorithm can be found in our earlier publications~\cite{HajiAkbariFilmMolinero2014, GianettiPCCP2016}. Forward flux sampling (FFS)~\cite{AllenFrenkel2006} has been extensively utilized for studying rare events~\cite{HussainJChemPhys2020}, and jFFS is a generalized variant of FFS particularly suited for use with order parameters-- such as the one utilized in this work-- that undergo high-frequency high-amplitude temporal fluctuations.
 The rate of transition from the supercooled  liquid basin, $A:=\{\textbf{x}:
\xi(\textbf{x})<\xi_A\}$ to the crystalline basin $B:=\{\textbf{x}:\xi(\textbf{x})\ge\xi_B\}$ is estimated by partitioning the intermediate $[{\color{black}\xi}_A,{\color{black}\xi}_B)$ region using $N$ milestones ${\color{black}\xi}_A<{\color{black}\xi}_0<\cdots<{\color{black}\xi}_N={\color{black}\xi}_B$, which are level sets of ${\color{black}\xi}(\cdot)$, and by recursively computing the flux of trajectories leaving $A$ and reaching each milestone. This is achieved by computing the flux of trajectories crossing ${\color{black}\xi}_0$ (computed from long conventional MD trajectories within $A$) and estimating transition probabilities between successive milestones (by initiating trial trajectories from configurations arising from earlier crossings).   
  The mechanism of nucleation is characterized using the pedigree analysis method described in Ref.~\citenum{HajiAkbariPNAS2015}. Further details about jFFS calculations can be found in  Section~\ref{SI-section:jFFS}.

\begin{figure}
\centering
\includegraphics[width=0.4913\textwidth]{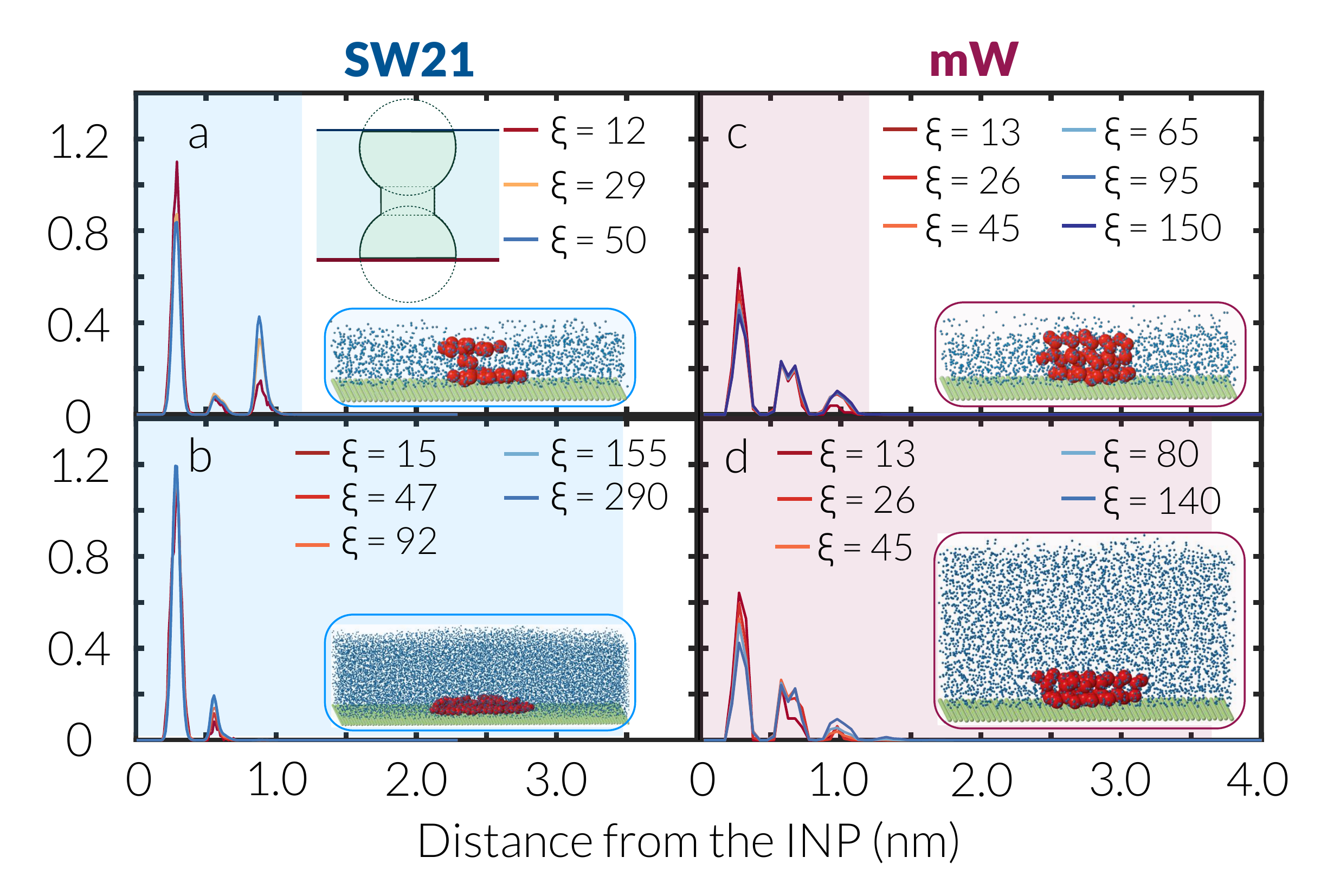} 
\caption{\textbf{Geometric Spread of Crystalline Nuclei in Graphene-supported Films.} Histograms of the \textit{z} coordinates of the molecules belonging to the largest crystalline nuclei in surviving configurations of SW21 (a-b) and mW (c-d) films at 170~K and 235~K, respectively. The area under each histogram is normalized to unity. Legends correspond to surviving nuclei at different jFFS milestones. Insets depict  representative critical nuclei in each system, with solid-like molecules within the nuclei, liquid-like molecules and carbon atoms depicted in red, dark blue and light green, respectively. Nuclei in (a) are hourglass-shaped with their typical geometry depicted in the inset.  Shaded regions correspond to the geometric spreads of the supported films. 
}
\label{Figure 2}
\vspace{-20pt}
\end{figure}

\section{Results and Discussion}\label{Results and Discussion}

\noindent\textbf{Kinetics and Mechanism of Nucleation:}
We first explore the dependence of heterogeneous nucleation rate on the thickness of the supported film, which is a measure of the proximity of the INP and the free interface. As depicted in Figure~\ref{Figure 1}e, the rate is virtually insensitive to film thickness in the mW system, which does not undergo surface freezing. Note that such lack of sensitivity is not an artifact of the relatively large nucleation rates at 235~K, and is also observed at 240~K, where nucleation rates are four orders of magnitude smaller.  In the SW21 system, which undergoes surface freezing, however, the rate is very sensitive to film thickness as can be seen in Figure~\ref{Figure 1}f, and is almost six orders of magnitude larger in the ultrathin 1.2-nm thick film than in thicker films.  These findings confirm our core hypothesis that there is a relationship between the ability of a free interface to enhance homogeneous and heterogeneous nucleation, as the contact freezing propensities of these two liquids match their respective surface freezing tendencies. 
We use the heuristics developed in our earlier publication~\cite{HussainJCP2020p} to show that these findings are not impacted by finite size effects, with the results of such analysis presented in Figure~\ref{SI-fig:finite-size-effects} and Table~\ref{tab:finite-size}. This is further confirmed by conducting MD simulations of the 1.2-nm film in a larger simulation box (Figure~\ref{SI-fig:larger-INP}) with the rate obtained from the  mean first passage time method\cite{WedekindJChemPhys2007} almost identical to that reported in Figure~\ref{Figure 1}f.  Also, these rates are tens of orders of magnitude larger than the homogeneous nucleation rates at identical temperatures (Table~\ref{tab:homogeneous-rates}).

In order to understand the origin of this contrasting behavior, we inspect the nucleation mechanism by analyzing the spatial spread of the largest crystalline nuclei in 'surviving` configurations. A configuration stored at an FFS milestone is called 'surviving` if it bears progeny at the target crystalline basin,~i.e.,~if at least one configuration at ${\color{black}\xi}_N$ can be traced back to it via a collection of trial trajectories. As can be seen in Figure~\ref{Figure 2}, a major qualitative difference is observed between the ultrathin SW21 film  and the remaining films. While the crystalline nuclei only form at the graphene surface in  mW films of all thicknesses (Figures~\ref{Figure 2}c-d and \ref{SI-fig:struct-intermediate}c-d) and in thicker SW21 films (Figures~\ref{Figure 2}b and \ref{SI-fig:struct-intermediate}a-b), they tend to emerge at either of the two interfaces in the ultrathin SW21 film (as can be seen from the representative surviving configurations and pathways depicted in Figure~\ref{SI-fig:geom-spread}) and grow to form hourglass-shaped nuclei. (Here, by "hourglass" we refer to a structure that is fatter at the top and the bottom than in the middle.) Representative critical nuclei depicted in the insets are also hourglass-shaped in the ultra-thin SW21 film (Figures~\ref{Figure 2}a, \ref{SI-fig:larger-INP} and \ref{SI-fig:geom-spread}) as opposed to spherical cap-like nuclei reminiscent of  classical heterogeneous nucleation  in  thicker films (Figure~\ref{Figure 2}b). In mW films, however, all such nuclei are spherical cap-like irrespective of film thickness, as depicted in Figures~\ref{Figure 2}c-d. In particular, nucleation exclusively starts at the graphene wall in ultrathin mW films (Figure~\ref{Figure 2}c), as demonstrated in the pathway depicted in Figure~\ref{SI-fig:pathway-mW}. 

Clearly, the dependence of mechanism on film thickness follows the same trend as that of the nucleation rate. Most notably, the dramatic enhancement in nucleation kinetics in ultrathin SW21 films is accompanied by an abrupt change in the shape and spatial spread of the crystalline nuclei. This change demonstrates a synergy between the two interfaces in the ultrathin film, which is likely responsible for faster nucleation presumably due to a decrease in the nucleation barrier. We use the FFS-MFPT method\cite{ThaparJCP2015} to compute nucleation barriers in SW21 films and observe that the nucleation barrier decreases abruptly upon decreasing the film thickness to 1.2~nm (Figure~\ref{Figure 3}a). We refer  to this effect as \emph{interfacial synergy} since the proximity of the two interfaces leads to increases in rate (and decreases in the nucleation barrier) that are otherwise  impossible in the presence of the isolated individual interfaces. We also observe further manifestation of this synergy by noting that surviving configurations at $\xi_0$ are more likely than vanishing configuration to possess strong free interfacial peaks as depicted in Figure~\ref{SI-fig:surviving-vanishing}a.

\begin{figure*}[htp]
\centering
\includegraphics[width=\textwidth]{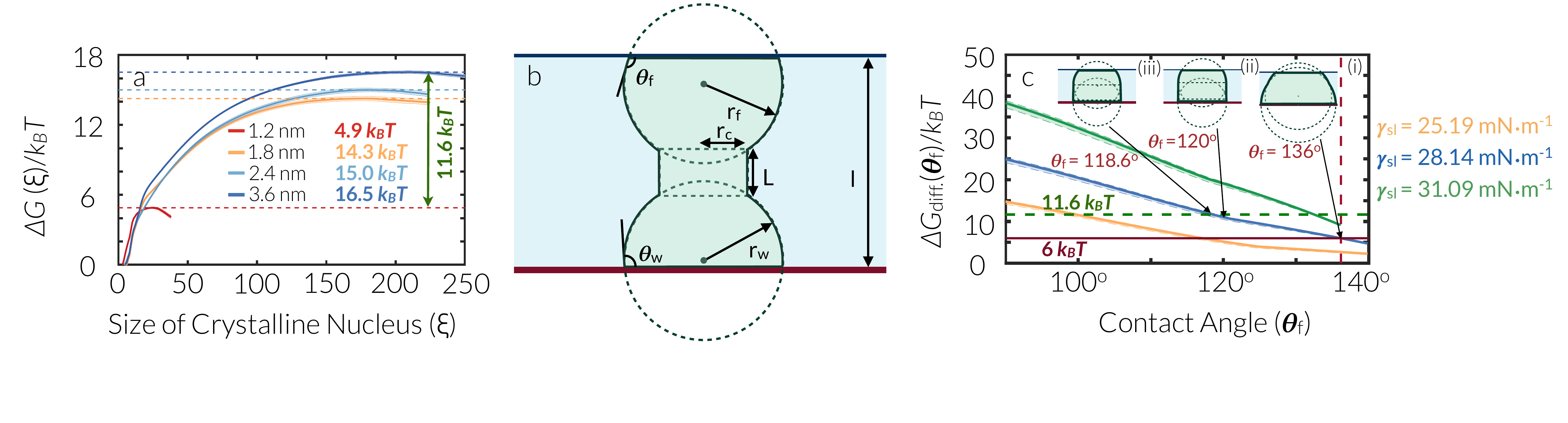} 
\vspace{-30pt}
\caption{\textbf{CNT-Based Theoretical Description of Nucleation:}  (a) Free energy profiles computed using the FFS-MFPT method~\cite{ThaparJCP2015} for nucleation within graphene-supported SW21 films. (b) Schematic representation of an hourglass-shaped crystalline nucleus. $\theta_\textit{w}$ and $\theta_\textit{f}$ are the corresponding contact angles at the INP and the free-interface, respectively, while $\textit{l}$ is the thickness of the film.  (c) $\Delta{\textit{G}}_{\text{diff}}$ vs.~$\theta_\textit{f}$ for the $\gamma_{\textit{sl}}$ given in Ref.~\citenum{GianettiPCCP2016} (blue). The orange and green curves are computed at the boundaries of $\gamma_{\textit{sl}}$'s 95\% confidence interval, while each dotted curve is computed at $\theta_\textit{w}\pm\delta\theta_\textit{w}$ with $\delta\theta_\textit{w}$ the error bar in $\theta_\textit{w}$.  The shade around each curve is therefore a measure of uncertainty in $\Delta{\textit{G}}_{\text{diff}}$ at a fixed $\gamma_{\textit{sl}}$ due to uncertainties in $\theta_\textit{w}$. The insets depict representative nucleus shapes predicted from the theory. The dark red and dark green horizontal lines correspond to the $\Delta{G}_{\text{diff}}$'s predicted from CNT for the uncorrected $\theta_f$ and estimated from the FFS-MFPT method, respectively.}
\vspace{-10pt}
\label{Figure 3} 
\end{figure*}

\noindent\textbf{CNT-Based Theoretical Model:}
In order to determine whether it is the formation of hourglass-shaped nuclei that results in a decrease in the nucleation barrier, we employ the formalism of classical nucleation theory~\cite{TurnballJCP1950}, which has been extensively utilized to interpret the findings of experimental and computational studies of nucleation~\cite{MichaelidesChemRev2016}. In the standard form of CNT for heterogeneous nucleation, crystalline nuclei are assumed to be spherical caps that form at the surface that harbors nucleation and grow at a fixed three-phase contact angle. This results in a nucleation barrier given by
\begin{eqnarray}\label{eq:barrier-het}
\Delta{G}_{\text{het}}^{*} &=& \frac{16\pi\gamma_{sl}^3f_c(\theta)}{3\rho_s^2|\Delta\mu|^2},
\end{eqnarray}
wherein $\Delta\mu$ is the chemical potential difference between the supercooled liquid ($l$) and the crystal ($s$), $\gamma_{sl}$ is the solid-liquid surface tension, $\rho_s$ is the number density of the crystal and  $\theta$ is the three-phase contact angle. $f_c(\theta)$-- given by Eq.~(\ref{eq:potency})-- is a measure of the efficiency of heterogeneous nucleation and is called the potency factor~\cite{CabrioluPRE2015}-- or the compatibility factor~\cite{AlpertPCCP2011}.
 We generalize this standard form of CNT to the case of two parallel  interfaces separated by distance $l$ by assuming that crystalline nuclei can comprise of two-- intersecting or non-intersecting-- spherical caps of radii $r_w$ and $r_f$ forming at the INP wall ($w$) and the free interface ($f$), respectively, and can be further connected via a cylindrical bridge of radius $r_c$. Within this framework, nucleation can start at both interfaces, commensurate with our observations in ultrathin SW21 films. Moreover, model nuclei can only touch each interface at a fixed contact angle ($\theta_w$ for the INP and $\theta_f$ for the free interface). This  further limits the set of permissible values of $r_w, r_f$ and $r_c$, as each cap can intersect the opposing interface only within the base of the opposing cap, and the cylindrical bridge will also have to be contained within those bases or not touch the interfaces at all. (See Section~\ref{SI-subsection:constraints} for a detailed discussion.)
The free energy of formation of such a composite nucleus (Figure~\ref{Figure 3}b) will thus be given by
\begin{eqnarray}
\Delta G_{\text{hg}}^l (r_w, r_f, r_c) &=& \sum_{i\in\{w,f\}} \pi r_i^2 \left (\gamma_{is} - \gamma_{il}\right)\sin^2{\theta_i}\notag\\&& -V_{\text{hg}} \rho_s|\Delta\mu| + \gamma_{sl} S_{\text{hg}}, \label{hg_energy}
\end{eqnarray}
with $V_{\text{hg}}$ and $S_{\text{hg}}$ the volume and the liquid-exposed surface area of the hourglass-shaped nucleus and $\gamma_{\alpha\beta}$ the surface tension between phases $\alpha$ and $\beta\in\{w, f, l, s\}$. The free energy of formation of a nucleus of size $N$ will thus be given by,
\begin{eqnarray}
\Delta{G}_{\text{hg}}^l(N) &=& \min_{\rho_sV_{\text{hg}}(r_w,r_f,r_c)=N} \Delta{G}_{\text{hg}}^l (r_w, r_f, r_c), \label{eq:GvsN}
\end{eqnarray}
The nucleation barrier $\Delta{G}_{\text{hg}}^{l,*}$ can be estimated by maximizing $\Delta{G}_{\text{hg}}^l(N)$. If the corresponding critical nucleus is comprised of a single spherical cap only,~i.e.,~with $r_w\neq0$ and $r_f=r_c=0$, $\Delta{G}_{\text{hg}}^{l,*}$ will be identical to $\Delta{G}_{\text{het}}^{*}$ given by Eq.~(\ref{eq:barrier-het}) and the proximity of two interfaces will not result in smaller barriers and faster nucleation. One can therefore use $\Delta{G}_{\text{diff}} = \Delta{G}_{\text{het}}^* - \Delta{G}_{\text{hg}}^{l,*}$ as a measure of the efficacy of the second interface in enhancing nucleation. 

Before discussing the predicted $\Delta{G}_{\text{diff}}$ values,
we first overview how we estimate the necessary thermodynamic parameters.
Unlike quantities such as $\Delta\mu$ and $\rho_s$ that can be accurately estimated via thermodynamic integration and $NpT$ MD simulations, surface tensions and contact angles are extremely difficult to estimate directly in the supercooled regime. For mW-like liquids, indirect estimates based on CNT reveal that $\gamma_{sl}$ is not very sensitive to temperature~\cite{VegaJCP2014} or tetrahedrality~\cite{GianettiPCCP2016}. We therefore use the value of $\gamma_{sl}=28.14 \pm 2.95$~mN$\cdot\text{m}^{-1}$ reported in Ref.~\citenum{GianettiPCCP2016}, which satisfactorily describes nucleation in mW-like liquids over a wide range of tetrahedralities. In order to estimate contact angles, we invoke a CNT prediction that has been previously validated in computational studies of heterogeneous ice nucleation on graphene~\cite{CabrioluPRE2015} and stipulates that  the potency factor is equal to the ratio of the sizes of critical nuclei in heterogeneous and homogeneous nucleation. We consider nucleation in freestanding films and supported 3.6-nm films as references for determining $\theta_f$ and $\theta_w$, respectively, with further details given in Section~\ref{SI-subsection:contact-angles}.We compute $\Delta{G}^l_{\text{hg}}(N)$ for SW21 films of different thicknesses (Figure~\ref{SI-fig:DeltaG-CNT}) using the numerical approach described in Section~\ref{SI-subsection:CNT:numerics} and observe that $\Delta{G}_{\text{hg}}^{l,*}$ is identical to $\Delta{G}_{\text{het}}^*$ in thicker films, and is only $6k_BT$ smaller in the ultrathin 1.2-nm film. The corresponding critical nucleus-- (i) in the inset of Figure~\ref{Figure 3}c-- is comprised of two intersecting spherical caps, as no cylindrical bridge is geometrically possible due to the sizes of the spherical caps.
These predictions are qualitatively consistent with the rates and mechanisms obtained from jFFS, and demonstrate that the synergy between the two proximal interfaces \emph{can} result in faster nucleation.  In mW films, however,  $\Delta{G}_{\text{hg}}^{l,*}=\Delta{G}_{\text{het}}^*$ for film of all thicknesses due to lack of surface freezing propensity at the free interface.

\noindent\textbf{Structural Characterization of the Free Interface in the Supercooled Liquid:}
Our theoretical description provides a qualitative explanation for faster nucleation in ultrathin SW21 films. Its quantitative accuracy, however, is limited as it underestimates the extent by which nucleation is enhanced,~i.e.,~a $6k_BT$ decline in barrier vs.~the $11.6k_BT$ obtained from the FFS-MFPT method and depicted in Figure~\ref{Figure 3}a. As can be seen from Table~\ref{tab:theory-errorbars}, such discrepancies cannot be fully explained by uncertainties in model parameters such as surface tensions and contact angles, thus they are  caused  by either limitations of CNT, or peculiarities specific to SW21 ultrathin films. In order to identify-- or rule out the existence of-- such  peculiarities, we analyze the molecular structure of the free interfaces in supported and freestanding liquid films of SW21 and mW. Here, ``free interface`` corresponds to parts of the film within the last major peak of $\rho(z)$, the density profile as a function of $z$, the distance from the wall and the film center for supported and freestanding films, respectively.  ($\rho(z)$ profiles for all the films considered in this work are depicted in Figure~\ref{SI-fig:density}.) We also compute each structural feature within the 'bulk` region, i.e.,~the parts of the films where density is constant and is equal to the bulk value,~e.g.,~at the center of freestanding films. We first compute $g(r,z)$, the planar radial distribution function (RDF)~\cite{HajiAkbariJCP2014}, which provides a radially-averaged picture of a molecule's hydration shells. As can be noted in Figure~\ref{Figure 4}a, there is a statistically significant difference between free interfacial RDFs of supported and freestanding ultrathin films of SW21, with the supported film RDF possessing a shallower first valley and a weaker second peak, corresponding to more intermixing of the first and second hydration  shells in the free interface. Moreover, the free interfacial RDF in the supported ultrathin film lies in between those in the bulk and the freestanding film. This suggests that the free interface becomes more bulk-like due to its proximity to the graphene wall. The distinction between the free interfacial RDFs of supported and freestanding films disappears in thicker SW21 films (Figures~\ref{Figure 4}b and \ref{SI-fig:struct-intermediate}e-f) and in mW films of all thicknesses (Figures~\ref{Figure 4}c-d and \ref{SI-fig:struct-intermediate}g-h).

In order to further probe the structure of the free interface, we compute the $q_3$ distribution for the molecules within the free interface. $q_3$ is a local Steinhardt  bond order parameter~\cite{SteinhardtPRB1983} usually used for distinguishing ice polymorphs~\cite{GalliPCCP2011}, but is, in general, a measure of how neighbors of a central molecule are oriented within its first hydration shell.  Similar to RDFs, free interfacial $q_3$ distributions differ considerably between supported and freestanding ultrathin SW21 films as depicted in Figure~\ref{Figure 4}e, while no such difference is observed in thicker SW21 films (Figures~\ref{Figure 4}f and \ref{SI-fig:struct-intermediate}{4}i-j) and in mW films of all thicknesses (Figures~\ref{Figure 4}g-h and \ref{SI-fig:struct-intermediate}k-l). The rightward shift in $q_3$ makes free interfaces in supported ultrathin films more bulk-like, a trend also observed for RDFs (Figure~\ref{Figure 4}a). Interestingly, this dramatic change in the $q_3$ distribution can be fully attributed to a change in the number of molecules within the first hydration shell, as depicted in Figure~\ref{Figure 4}i. Indeed, the $q_3$ distribution undergoes a rightward shift when the number of molecules within the first hydration shell increases (Figure~\ref{SI-fig:q3-ngb}). Again, no change in the nearest neighbor count distribution is observed in thicker SW21 films (Figures~\ref{Figure 4}j and  \ref{SI-fig:struct-intermediate}m-n) and mW films of all thicknesses (Figures~\ref{Figure 4}k-l and \ref{SI-fig:struct-intermediate}o-p). One might expect the synergy between the two interfaces to also result in a change in the structure of the graphene-adjacent  interfacial region, i.e.,~the region corresponding to the first density peak in Figure~\ref{SI-fig:density}a. Our analysis, however, reveals no such structural modulation, as evident from planar RDFs, and $q_3$  and nearest neighbor count distributions depicted in  Figure~\ref{SI-fig:INP-adjacent}. 

\begin{figure}[ht]
\centering
\includegraphics[width=0.4394\textwidth]{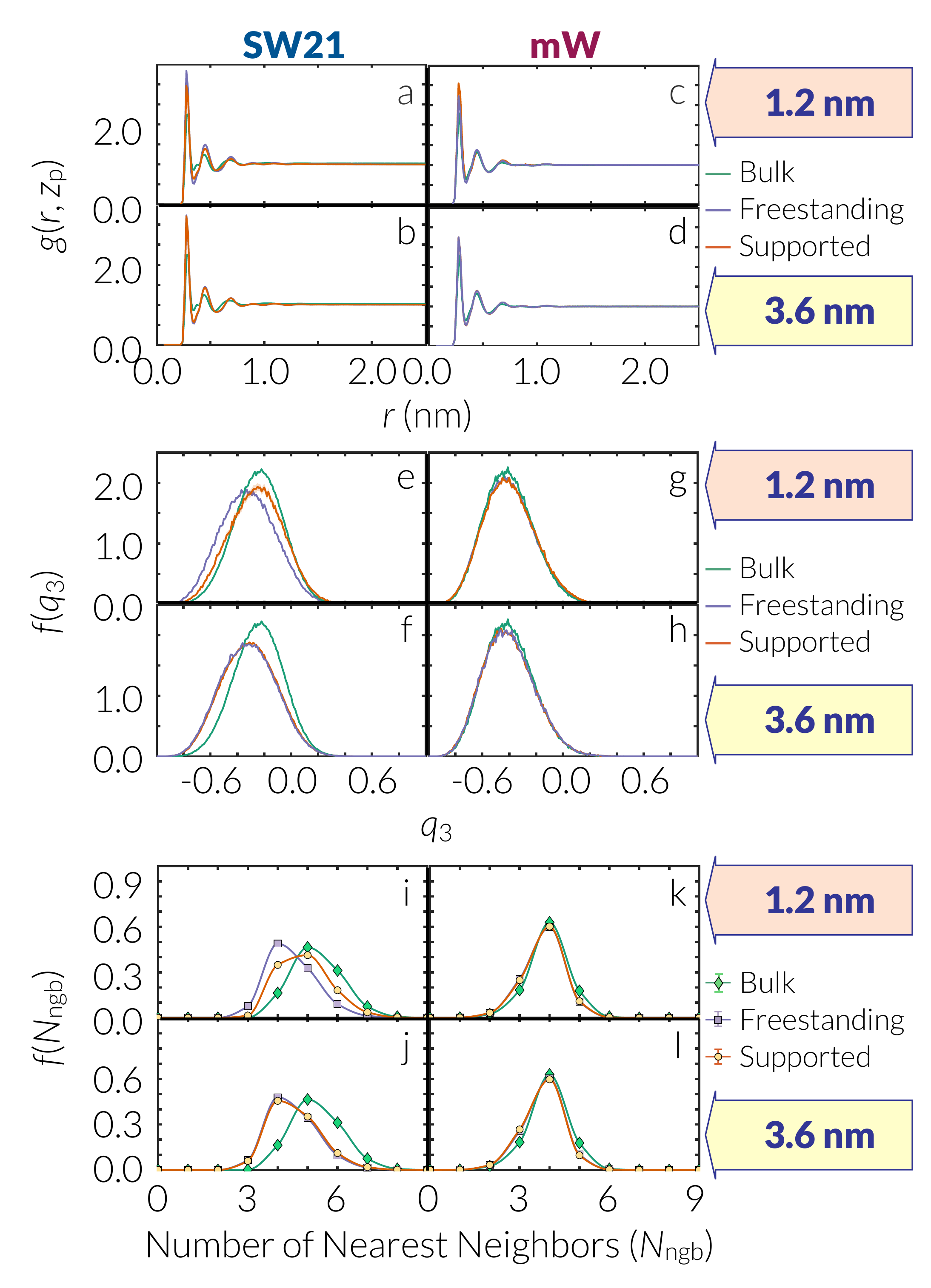} 
\caption{\textbf{Structural Characterization of the Free Interface:} (a-d) Planar RDF's, and (e-h) $\textit{q}_\text{3}$ and (i-l) nearest neighbor count distributions for molecules within the free interfacial regions of supported and freestanding thin films of SW21 (a-b, e-f, i-j) and mW (c-d, g-h, k-l). The ``bulk`` refers to the properties computed in the central bulk-like region of a 3.6-nm-thick freestanding film. Error bars in (a-d) and (i-l) are thinner than the curves and smaller than the symbols, respectively. The areas under the curves in (e-h) are normalized to unity, and the shades correspond to error bars.}
\label{Figure 4}
\vspace{-10pt}
\end{figure} 

\begin{figure*}[htp]
\centering
\includegraphics[width=.8117\textwidth]{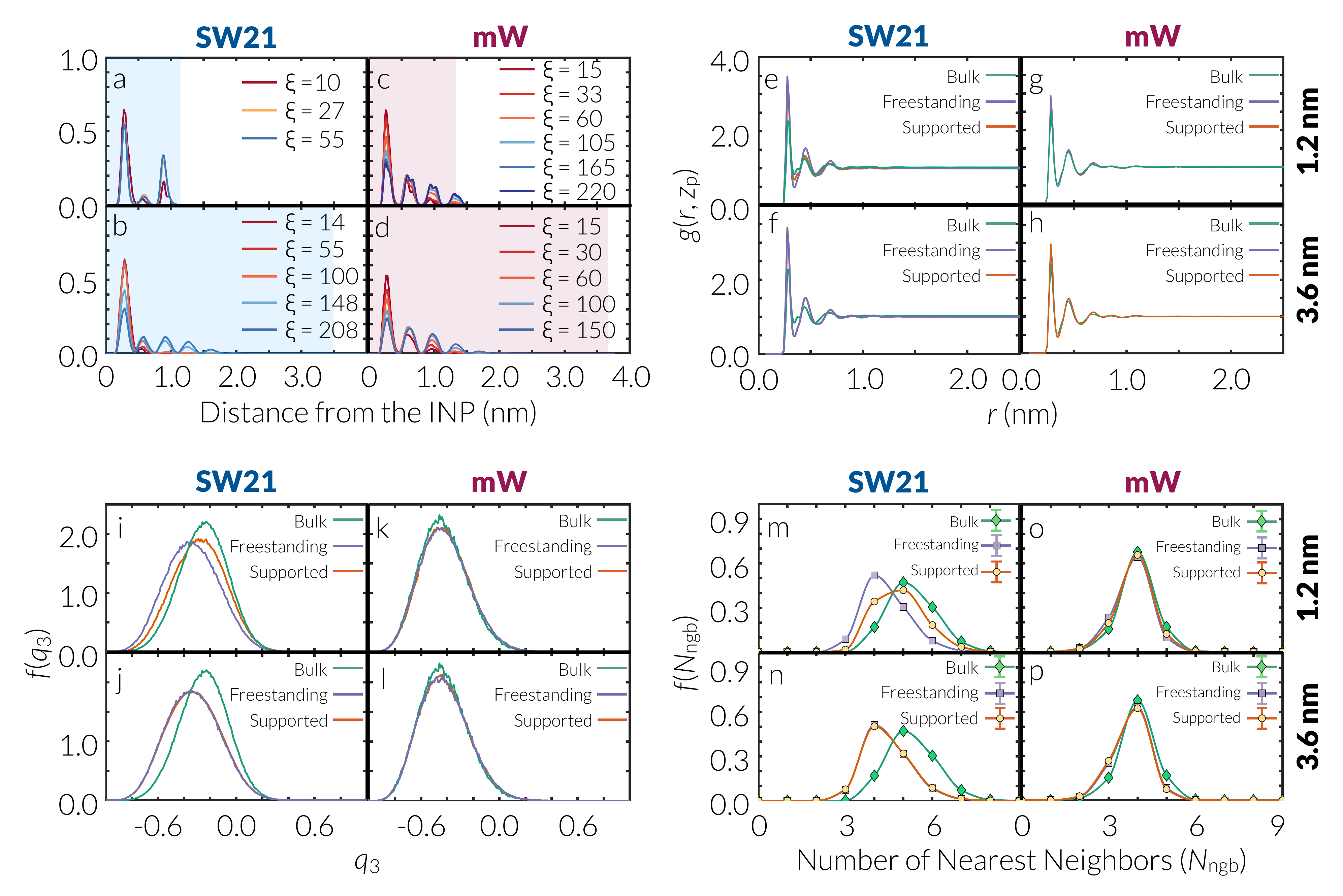} 
\caption{\textbf{Nucleation Mechanism and Structural Characterization in Supported films in the Vicinity of a Structureless INP:} (a-d) Histograms of the $\textit{z}$ coordinates of the molecules belonging to the largest crystalline nuclei in  surviving configurations of SW21 (a-b) and mW (c-d) films. Shaded regions correspond to the geometric spreads of the supported films. (e-h) Planar RDF's, and (i-l) $\textit{q}_\text{3}$ and (m-p) nearest neighbor count distributions for the vapor-liquid interfacial regions of supercooled supported and freestanding films of SW21 and mW. The "bulk`` refers to the properties computed in the central bulk-like region of a 3.6-nm-thick freestanding film. Error bars in (e-h) and (m-p) are thinner than the curves and smaller than the symbols, respectively. The areas under the curves in (i-l) are normalized to unity and the shades correspond to error bars.}
\label{Figure 5}
\vspace{-10pt}
\end{figure*}

All these structural features point to the same picture, a free interface that becomes increasingly bulk-like in the presence of a proximal INP. Such changes in structure will inevitably alter interfacial properties such as $\gamma_{lv}$ and $\theta_f$. In particular, we expect $\gamma_{lv}$ to only increase upon an INP-induced structural modulation, since the unperturbed free interface adopts the structure that minimizes the free energetic penalty associated with forming a two-phase interface. Any deviation from such 'optimal` structure will only increase such penalty. According to the Young equation, $\theta_f$ is related to $\gamma_{lv}$ by $\cos\theta_f=(\gamma_{lv}-\gamma_{sv})/\gamma_{sl}$, and will therefore decrease upon an increase in $\gamma_{lv}$. This is based on the reasonable assumption that $\gamma_{sv}$ is not affected by the thinness of the film. Unfortunately, we cannot accurately estimate the perturbed $\gamma_{lv}$ using standard methods such as integrating the difference between normal and lateral stress and the capillary wave method~\cite{IsmailJCP2006} due to the absence of a well-defined bulk region, nor can we compute it using the test area method~\cite{GloorJCP2005}, which will require straining the crystalline graphene wall. We therefore only examine the sensitivity of $\Delta{G}_{\text{diff}}$ to $\theta_f$. As depicted in Figure~\ref{Figure 3}c, $\Delta{G}_{\text{diff}}$ increases upon decreasing $\theta_f$ from its unperturbed value of $\approx136^{\circ}$. Indeed, decreasing $\theta_f$ by $\approx18^\circ$ brings $\Delta G_{\text{diff}}$ up to the $11.6~k_B T$, depicted in Figure~\ref{Figure 3}a. Moreover, the ''perturbed`` value of $\theta_f\approx118.6^{\circ}$ results in critical nucleus shapes and sizes more commensurate with those obtained from jFFS. For instance, unlike the original nucleus (predicted for the original $\theta_f\approx136^{\circ}$) that is only comprised of two spherical caps, all critical nuclei for  $\theta_f\le120.6^\circ$ also comprise a cylindrical bridge, and therefore more resemble the hourglass-shaped nuclei observed in simulations.  Furthermore, with the "corrected`` $\theta_f\approx118.6^\circ$, the critical nucleus size is predicted to be $N^*_{\text{corr}}\approx150{\color{black}^{+146}_{\color{black}-75}}$, which is considerably smaller than  $N^*\approx216{\color{black}^{+113}_{-79}}$ (predicted for $\theta_f\approx136^\circ$), and is closer to the average critical nucleus size of $N^*_{\text{jFFS}}\approx{\color{black}31\pm2}$ obtained from jFFS. Note that the  difference between $N^*_{\text{corr}}$ and $N^*_{\text{jFFS}}$ might be inflated as the apparent nucleus size determined from classical MD or jFFS is usually very sensitive to the employed clustering and classification algorithm~\cite{LupiNature2017}. Indeed, analyzing our critical configurations using a second order parameter in which the first hydration shells of solid-like molecules are included in the nucleus yields an average nucleus size of {\color{black}92$\pm$1}, which is considerably closer to $N^*_{\text{corr}}$ and falls within its confidence interval. Finally, as can be seen in Figure~\ref{Figure 3}c, our analysis is robust to uncertainties in the model parameters such as $\gamma_{sl}$ even though the 'corrected` $\theta_f$ that would yield the expected $\Delta G_{\text{diff}}$ will be slightly different. The confidence intervals for the 'corrected` $\theta_f$ and its associated $N^*$ is given in Table~\ref{tab:theory-errorbars}.

It must be noted that the quantitative accuracy of our CNT-based model is still limited even with an adjusted $\theta_f$. Most importantly, the predicted nucleation barriers (for both thick and thin films) are considerably larger than those estimated from the FFS-MFPT method.  This discrepancy can arise from, among other things, the strong sensitivity of the nucleation barrier to  quantities such as surface tensions and contact angles. For instance, a 10\% error in $\gamma_{sl}$ and $\theta_w$ can result in as much as 33\% and 25\% error in $\Delta{G}^*_{\text{het}}$, respectively. It is indeed plausible that we might be overestimating $\theta_w$, as the typical critical nuclei on a 3.6-nm SW21 film (e.g.,~the one depicted in Figure~\ref{Figure 2}b) are too flat to be approximated as spherical caps.  Another factor that can impact our contact angle estimates is the classification and clustering criteria utilized for detecting the largest crystalline nucleus, which can result in large changes in the apparent size of the critical nucleus~\cite{LupiNature2017}.  Despite these limitations, our analysis is still useful as it demonstrates that the synergy between interfaces and structural modulation of the free interface by an INP  can collectively explain the observed acceleration of nucleation in ultrathin SW21 films.

Our structural analysis of the free interface also sheds further light into the unresolved conundrum  of why different water models have such distinct surface freezing propensities. For all the structural features highlighted in Figure~\ref{Figure 4}, the interfacial regions in freestanding SW21 films are  distinct from the bulk, while for mW films, no difference is observed between the bulk and the free interface. This significant difference between these two otherwise similar models can qualitatively explain their differing surface freezing propensities. In other words, in order for the free interface to harbor nucleation at a faster rate, its structure must be sufficiently different from the bulk.

\noindent\textbf{Structureless Walls:}
In order to assure that the observed behavior is not an artifact of the molecular structure of the underlying graphene wall and is truly caused by a synergy between a free interface and a ''generic`` INP,  we explore the kinetics and mechanism of nucleation in thin supported SW21 and mW films in the vicinity of LJ 9-3 structureless walls~\cite{MagdaJCP1985}. A structureless wall exerts no lateral force on the molecules and is therefore incapable of inducing any lateral order within the film. Table~\ref{table:structureless} summarizes the computed nucleation rates in 1.2-nm and 3.6-nm thick supported mW and SW21 films. In order to keep the corresponding calculations computationally tractable and devoid of finite size effects, we conduct them at slightly lower temperatures.  Similar to graphene-supported films, the nucleation rate is virtually insensitive to film thickness in the mW system. In the SW21 system, however, nucleation is 16 orders of magnitude faster in the ultrathin film. These findings confirm that even though the extent by which nucleation becomes faster can depend on the particular structure and chemistry of the INP, the mere enhancement in heterogeneous nucleation kinetics only depends on the surface freezing propensity of the corresponding liquid.

\begin{table}
\caption{Summary of heterogenous nucleation rate, $\mathcal{J}$, in supported SW21 and mW films in the vicinity of an LJ 9-3 structureless INP. Error bars correspond to 95\% confidence intervals.}
\centering
\begin{tabular}{p{1.3 cm}p{1.3cm}p{2.1 cm} p{2.1 cm}}
\hline\hline
\multirow{2}{*}{Model} & \multirow{2}{*}{T~(K)}& \multicolumn{2}{c}{$\text{log}_{\text{10}}\mathcal{J}~[\text{m}^{-2}\cdot\text{s}^{-1}]$} \\
\cline{3-4}
& & $1.2$-nm fim  & $3.6$-nm film \\
\hline
SW21 & 165 & $22.96\pm0.26$  & $6.82\pm0.45$  \\
\hline
mW & 220 & $20.07\pm0.08$ & $20.86\pm0.07$ \\
\hline
\end{tabular}
\vspace{-5pt}
\label{table:structureless}
\end{table}

\begin{figure}[ht]
	\centering
	\includegraphics[width=.3665\textwidth]{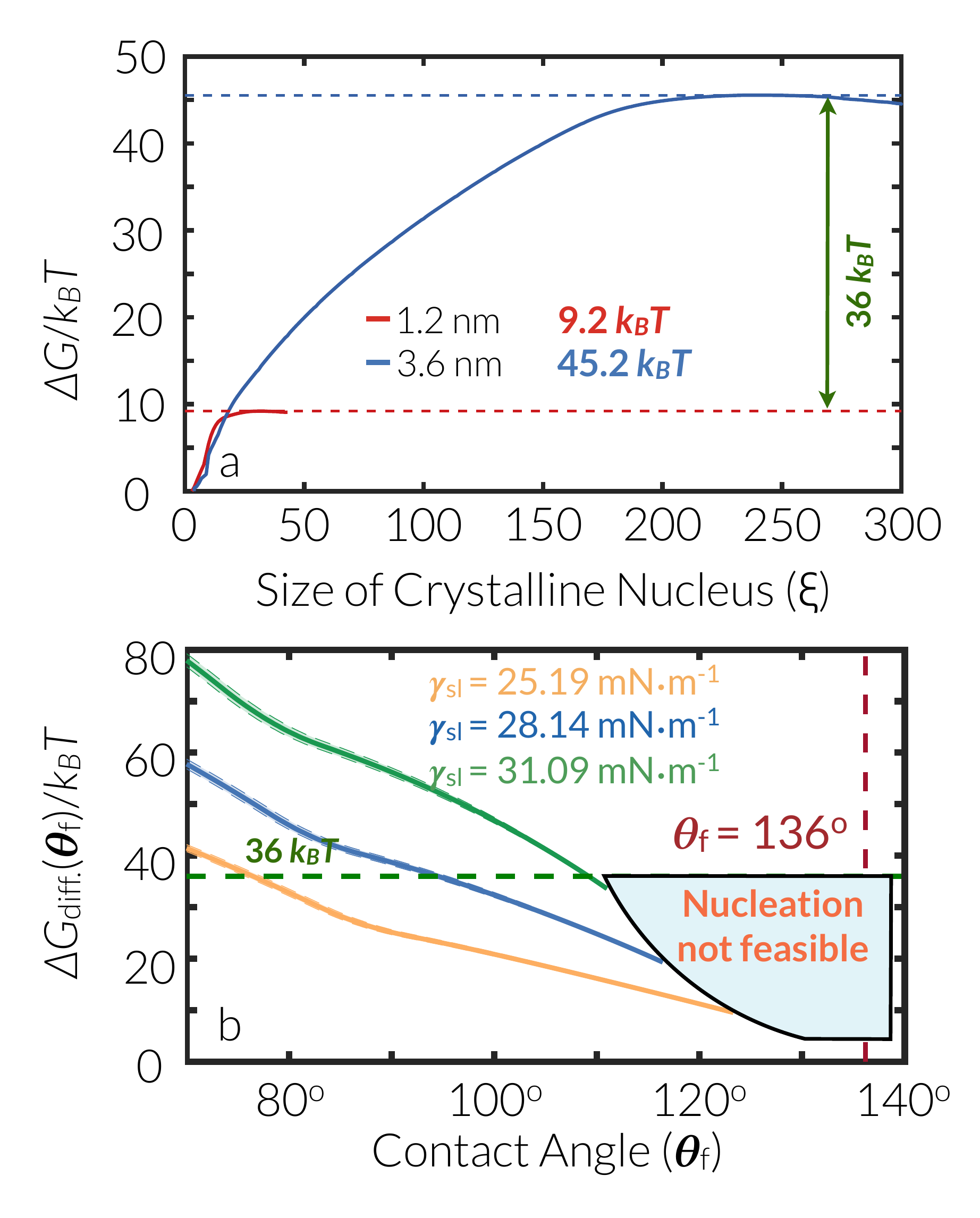}
	\vspace{-10pt}
	\caption{{\textbf{Theoretical Analysis of Nucleation in the Vicinity of a Structureless INP:} (a) Free energy profiles computed using the FFS-MFPT method~\cite{ThaparJCP2015} for nucleation within supported SW21 films in the vicinity of the structureless INP. (b)} $\Delta\textit{G}_{\text{diff}}$ vs.~$\theta_\text{f}$ for the ultrathin film at the vicinity of the structureless wall for the $\gamma_{\textit{sl}}$ given in Ref.~\citenum{GianettiPCCP2016} (blue). The orange and green curves are computed at the boundaries of $\gamma_{\textit{sl}}$'s 95\% confidence interval, while each dotted curve is computed at $\theta_\textit{w}\pm\delta\theta_\textit{w}$ with $\delta\theta_\textit{w}$ the error bar in $\theta_\textit{w}$. The shade around each curve is therefore a measure of uncertainty in $\Delta{G}_{\text{diff}}$ at a fixed $\gamma_{\textit{sl}}$ due to uncertainties in $\theta_\textit{w}$. Nucleation is not feasible within the light blue region. The dark green horizontal lines corresponds to the $\Delta{G}_{\text{diff}}$ estimated from the FFS-MFPT method in (a).}
	\label{Figure 6}
	\vspace{-10pt}
\end{figure}

We also explore the mechanism of nucleation by quantifying the spatial spread of crystalline nuclei in surviving configurations. In ultrathin SW21 films, crystalline nuclei form at both interfaces and are hourglass-shaped (Figure~\ref{Figure 5}a), a behavior also observed in their graphene-supported counterpart (Figure~\ref{Figure 2}a). Similarly, nucleation proceeds through conventional heterogeneous nucleation in thicker SW21 films (Figure~\ref{Figure 5}b) and in mW films of all thicknesses (Figures~\ref{Figure 5}c-d). Moreover, structureless walls modulate the structure of the free interface in ultrathin SW21 films (Figures~\ref{Figure 5}e,i,m) while no such modulation is observed in thicker SW21 films (Figures~\ref{Figure 5}f,j,n) and mW films of all thickness (Figures~\ref{Figure 5}g-h,k-l,o-p).

Applying our CNT-based theory to nucleation near structureless walls is particularly instructive. Since both $\theta_w$ and $\theta_f$ are obtuse in the case of the structureless wall, a geometric upper bound exists for the sizes of nuclei of the type depicted in Figure~\ref{Figure 3}b. (A rigorous proof is provided in Section~\ref{SI-subsection:obtuse}.) For the unperturbed $\theta_f$, $\Delta{G}_{\text{hg}}^l(N)$ is a strictly increasing function of $N$, and therefore no nucleation is feasible according to the theory. The structural modulation of the free interface, however, implies that $\theta_f$ is smaller in supported ultrathin SW21 films. Decreasing $\theta_f$ not only makes nucleation possible, but also results in a larger $\Delta{G}_{\text{diff}}$ (Figure~\ref{Figure 6}b). Using a perturbed contact angle of $\approx 94.6^{\circ}$ yields  $\Delta{G}_{\text{diff}}\approx36~k_BT$ obtained from the FFS-MFPT method and depicted in Figure~\ref{Figure 6}a.
 Similar to graphene walls, this analysis is robust to uncertainties in model parameters such as $\gamma_{sl}$ and $\theta_w$ as can be seen in Table~\ref{tab:theory-errorbars}.

\noindent\textbf{Polymorphism and Cubicity of Crystalline Nuclei:}
Figure~\ref{Figure 7}  depicts the percentage of molecules within the largest crystalline nuclei with local structure of cubic ice, determined using the $q_3$ order parameter~\cite{GalliPCCP2011}. It is abundantly clear that both INPs favor the formation of hexagonal ice at their immediate vicinity. Consequently, cubicity is very small in the case of graphene-supported SW21 films (Figure~\ref{Figure 7}a) since the crystalline nuclei are predominantly comprised of two layers in thicker SW21 films (Figures~\ref{Figure 2}b and \ref{SI-fig:struct-intermediate}a-b).  In the 3.6-nm SW21 film supported by the structureless INP, cubicity is higher as  the crystalline nuclei are comprised of more layers than their graphene-supported counterparts (Figure~\ref{Figure 5}b). As for the ultrathin SW21 films, cubicity is very small since both interfaces tend to favor the formation of hexagonal ice. It has indeed been previously shown that free interfaces tend to favor the formation of hexagonal ice motifs in a wide variety of water models, including mW~\cite{HudaitJACS2016, HajiAkbariPNAS2017}. Unlike SW21 films, supported mW films are generally more cubic (Figure~\ref{Figure 7}c-d). While this can be partly attributed to the existence of more layers within the crystalline nuclei (Figures~\ref{Figure 2}c-d, \ref{Figure 5}c-d  and \ref{SI-fig:struct-intermediate}c-d), it can also be due to the higher propensity of mW towards stacking disorder. Further studies are needed to assess the sensitivity of polymorphism to the tetrahedrality parameter in the mW-like systems.

\begin{figure}[htp]
	\centering
	\vspace{-10pt}
	\includegraphics[width=.5\textwidth]{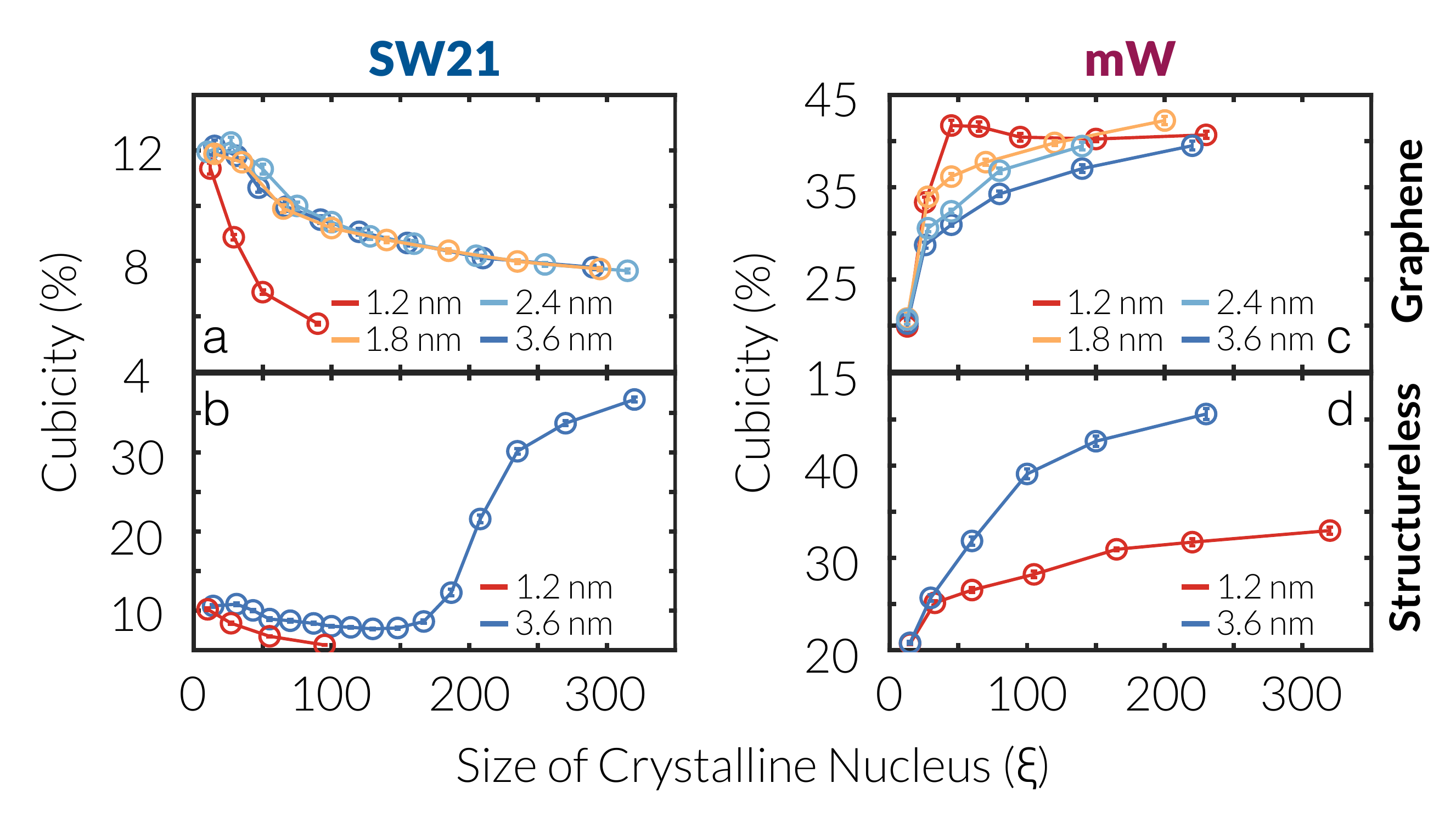}
	\vspace{-10pt}
	\caption{\textbf{Cubic Content of Crystalline Nuclei:} Percentage of solid-like molecules within the crystalline nuclei with the local structure of cubic ice in (a-b) SW21 and (c-d) mW films in the vicinity of (a,c) the graphene and (b,d) the structureless INP. }
	\label{Figure 7}
\end{figure}

\noindent\textbf{Comparison with Experiments:} Due to the limited spatiotemporal resolution of the existing experimental techniques, the mechanistic details obtained here cannot be directly verified in experiments. We can, however, compare the kinetic enhancements observed here to those reported in contact freezing experiments. Due to the interfacial nature of heterogeneous nucleation, we report all rates in nucleation events per unit area per unit time.  This is in contrast to experiments where an average volumetric nucleation rate is reported for an ensemble of microdroplets. It is therefore necessary to build a kinetic model to convert our areal rates (computed in idealized systems) to the apparent volumetric rates measured in experiments.  More specifically, $\mathcal{J}_a(s,\mathbf{\Phi})$, the areal rate of heterogeneous nucleation on an INP will depend on $s$, its distance from the free interface, and $\mathbf{\Phi}$, its orientation relative to the free interface. The average volumetric nucleation rate for a droplet of radius $r_0$ with $n$ dispersed INPs will thus be given by,
\begin{eqnarray}
\mathcal{J}_v[r_0;p(\cdot)]=\frac{3an}{r_0^3}\int_0^{r_0}\int_{\mathbf{\Phi}} r^2p(r,\overline{\mathbf{\Phi}})\mathcal{J}_a(r_0-r,\overline{\mathbf{\Phi}})  d\overline{\mathbf{\Phi}}dr,\notag\\
\label{eq:Jv-general}
\end{eqnarray}
where $a$ is the  surface area of an individual INP and $4\pi r^2p(r,\mathbf{\Phi})dr d\mathbf{\Phi}$ is the probability of observing it at a distance $r$ from the center and at a relative orientation $\mathbf{\Phi}$. Both $\mathcal{J}_a(\cdot)$ and $p(\cdot)$ can, in principle, be constructed using a combination of thermodynamic analysis and extensive molecular simulations. Note that Eq.~(\ref{eq:Jv-general}) is only valid if the INPs do not 'interact` with one another, i.e.,~that their nanoscale proximity and/or aggregation does not result in faster nucleation. 

The simple physical picture emerging from this work suggests that $\mathcal{J}_a$ can take two distinct values $\mathcal{J}_a^{(s)}$ and $\mathcal{J}_a^{(i)}$ for $s\le s_0$ and $s>s_0$, respectively, where $s_0$ is the threshold for transitioning from hourglass-shaped to regular nuclei and $\mathcal{J}_a^{(s)}\gg\mathcal{J}_a^{(i)}$. We also expect the interfacial contribution to Eq.~(\ref{eq:Jv-general}) to be dominated by the orientation in which the INP is parallel to the free interface. This is because an arrangement in which the two interfaces are proximal, but make a non-zero angle, is not mechanically stable, and depending on the wetting properties of the INP will either  revert to the parallel arrangement or will partially de-wet (and form a contact line) over timescales considerably shorter than the nucleation time. (We do not expect the latter scenario to result in considerable changes in $\mathcal{J}_a^{(s)}$ as contact lines have been shown to not  accelerate heterogeneous ice nucleation on chemically uniform surfaces~\cite{GurganusJPhysChemLett2011, GurganusJPhysChemC2013}.)
 The only plausible reason for $\mathcal{J}_a^{(s)}$ to be orientation-dependent is if the underlying INP has different crystallographic planes with differing ice-nucleating potencies. Depending on which of these planes comes in contact with the free interface, the extent by which nucleation is enhanced might be different. Considering the simple geometries of the INPs considered in this work (a single layer of graphene, and a structureless INP with no crystallographic features), this latter situation is not relevant here. 
If the INPs are also uniformly distributed within the droplet, the apparent volumetric rate will be given by:
\begin{eqnarray}
\mathcal{J}_v^{(u)} &=& an\mathcal{J}_a^{(i)}(3\alpha\varsigma+1), \label{eq:Jv-uniform}
\end{eqnarray}
with $\alpha=s_0/r_0\ll1$ and $\varsigma=\mathcal{J}_a^{(s)}/\mathcal{J}_a^{(i)}\gg1$. (The derivation of Eq.~(\ref{eq:Jv-uniform}) is included in Section~\ref{SI-section:derivation}.) We call $3\alpha\varsigma$ the \emph{enhancement factor} as it is the factor by which the apparent volumetric rate is enhanced due to "inside-out" freezing, and denote it by $\chi$. According to the calculations conducted here, $s_0$ is in the order of a few nanometers, while $\varsigma$ varies between $10^6$ to $10^{16}$. This will correspond to $3\alpha\varsigma\sim10^3-10^{13}$ for a microdroplet, which is in line with the enhancements observed in earlier experimental studies~\cite{ShawJPCB2005, ForneaJGeophysRes2009} that report enhancements between 5 to 13 orders of magnitude. 

While this simple kinetic model predicts $\chi$'s that are in reasonable agreement with experiments, it is important to assess its robustness to violations of some of its key underlying assumption. In particular, we consider a situation in which $\mathcal{J}_a(s)$ is not constant within the interfacial region ($s\le s_0$). It must, however, be noted that even then $\mathcal{J}_a(s)$ is unlikely to be an arbitrary continuous function of $s$. This is because a liquid nanofilm that lies in between an INP and the free interface will be layered as can be seen in Figure~\ref{SI-fig:density}. Consequently, not only the thickness of such a film will change in increments of  $0.3-0.4$~nm (the characteristic thickness of each liquid layer), but also  $\mathcal{J}_a(s)$ will be a discontinuous function of $s$ and will only depend on the number of liquid layers that separate the INP and the free interface. 
As demonstrated in detail in Section~\ref{SI-section:derivation}, the enhancement factor obtained from such a stratified model will be dominated by contributions from the "magic`` separations at which $\mathcal{J}_a(s)/\mathcal{J}_a^{(i)}$ is the largest. This will make our predictions robust to a "worst-case`` scenario in which nucleation is only enhanced within a 1.2-nm film (i.e.,~one comprised of three full liquid layers) and not for films that are thicker or thinner. Under such a scenario, $\mathcal{J}_v(r_0)$ will be dominated by contributions from $s\approx1.2$~nm, which will eclipse contributions from slow nucleation at other permissible thicknesses. 
As a result, the  enhancement factor will only decrease by an algebraic factor given by Eq.~(\ref{eq:modified-enhancement-factor}) while its order of magnitude will remain unchanged.  Therefore, the enhancement in rate within a 1.2-nm film alone is still sufficient for a dramatic increase in $\mathcal{J}_v$ even if it ditsappears for films with fewer or more liquid layers.

\section{Conclusion}\label{Conclusion}

\noindent
In this work, we explore how free interfaces impact heterogeneous ice nucleation by computing heterogeneous nucleation rates in supported supercooled nanofilms of two model water-like tetrahedral liquids. We observe that the kinetics of nucleation is enhanced by several orders of magnitude in ultrathin films of the liquid that undergoes surface freezing, i.e.,~that has a free interface amenable to homogeneous nucleation. No such enhancement is observed for the liquid with no surface freezing propensity.  We use classical nucleation theory to conclude that the formation of hourglass-shaped crystalline nuclei (observed in our jFFS simulations of the films that undergo faster nucleation) can result in a considerable decrease in the nucleation barrier, but not by enough to explain the extent of increase in rate. By analyzing the structure of the supercooled liquid, we observe that the INP alters the structure of the free interface in the ultrathin films that undergo faster heterogeneous nucleation, and makes it more bulk-like. This results in a decrease in the three-phase contact angle at the free interface, which in turn leads to smaller nucleation barriers and faster nucleation. We confirm these findings for both graphene and model structureless LJ 9-3 walls.

Both model INPs considered in this work induce significant structural perturbations within the free interfacial region of the ultrathin SW21 film, while the INP-adjacent interface is mostly unaffected by the free interface. Note that either of these assertions might be violated for INPs with differing topographies and chemistries. As demonstrated in the case of graphene, faster nucleation can still be possible in the absence of INP-induced structural modulations, but the extent of enhancement will  be attenuated considerably. Further studies with a wide variety of INPs are needed to probe whether and when any of these key observations are violated.  

Our work provides ample evidence that  nanoscale proximity of an INP and a vapor-liquid interface can lead to rate increases commensurate with those observed in contact nucleation. 
There are, however, reasons to suspect that these findings might have limited direct relevance to atmospheric contact freezing, which occurs under conditions far from equilibrium, and is likely impacted by a plethora of other factors. Therefore, even though we demonstrate that nanoscale proximity is a sufficient condition for kinetic enhancement in contact nucleation, it is plausible that the inclusion of those other effects might result in comparable (or even larger) increases in the nucleation rate. Further studies are needed to assess the relative importance of factors such as etching\cite{FletcherJAtmosSci1970}, vapor deposition\cite{CooperJAtmosSci1974} and mechanical waves\cite{FukutaJAtmosSci1975}.  Moreover, the validity of the physical picture presented here is predicated on the assumption that real water undergoes surface freezing, which, while supported by a large body of indirect evidence, is yet to be proven unequivocally~\cite{HajiAkbariJCP2017}.

Recently, pressure perturbations have been proposed~\cite{YangAtmosphere2020} as a plausible cause of kinetic enhancement during contact nucleation. According to this theory, a collision between an INP and a water droplet could result in the formation of a distorted contact line, and thus lead to the emergence of regions with local negative curvature. The ensuing negative Laplace pressure will then result in faster nucleation due to water's negatively sloped melting curve~\cite{MarcolliSciRep2017}. While this theory cannot be fully confirmed experimentally due to the difficulties of probing nanoscale local curvature, it tends to perform reasonably well in explaining experimental observations of contact nucleation efficacy. While regions with local negative curvature can arise in our simulations, e.g.,~due to capillary waves at the free interface, the flat geometry of the supported film makes it extremely unlikely for such regions to extend over sufficiently large swaths of the liquid. Therefore, our work reveals that the emergence of negative pressure is not a necessary condition for faster contact nucleation. Further studies are, however, necessary to probe the combined effect of interfacial curvature and nanoscale proximity on the kinetics and mechanism of heterogeneous nucleation.

Due to the coarse-grained nature of the utilized force-fields, our explanation for the relationship between surface and contact freezing is minimal in nature. In particular, we are not able to capture electrostatic and polarizability effects that play an important role in heterogeneous ice nucleation, as demonstrated in several earlier studies~\cite{YanJPhysChemLett2011, FrauxJCP2014,  SossoJPhysChemLett2016, GlatzJChemPhys2016}. Due to the long-range nature of electrostatic interactions, the synergy between an INP with charged or polar groups and the free interface might be stronger and might extend over longer distances. It is therefore likely that the enhancement in nucleation kinetics will occur for films that are considerably thicker than a nanometer. The nature of INP-induced structural modulations might also be different as the free interface in real water has distinct dielectric signatures due to the presence of dangling hydrogen bonds~\cite{DuPhysRevLett1993}. Exploring these questions can be the topic of future studies.

Despite the success of our CNT-based theory in predicting faster nucleation in ultrathin SW21 films, its predictive power at a quantitative level is limited due to a confluence of factors, such as the difficulty in accurately estimating interfacial properties such as surface tensions and contact angles and the exponential sensitivity of rate to subtle changes in such quantities. The more important-- and consequential-- shortcoming of CNT, however, arises from the important role of INP-induced structural modulation that effectively alters the relevant interfacial properties. Therefore, even if all interfacial properties are estimated accurately, and even if CNT accurately describes both homogeneous and heterogeneous nucleation, it will still fall short of accurately predicting the extent by which contact freezing will be faster. This partly explains the quantitative inadequacy of  CNT-based models in describing contact freezing in experiments.

We wish to conclude with a few broader implications of this work beyond contact nucleation. First of all, our findings call for a more cautious approach in interpreting immersion nucleation experiments in which a large number of water microdroplets are generated from a mixture of water and INP particles. The fraction of the microdroplets that freeze upon supercooling is then monitored as a function of time, and an average nucleation rate is extracted accordingly~\cite{HajiAkbariJCP2017}. It is totally plausible that the INPs within such droplets might approach the free interfacial region and harbor nucleation at considerably larger rates in accordance with the mechanism discovered in this work. The emergence of such ''nucleation hotspots`` can, in turn, result in an overestimation of the true immersion nucleation rate, as suggested by Eqs.~(\ref{eq:Jv-general}) and (\ref{eq:Jv-uniform}). Such nanoscale proximity will be more likely to emerge if the INPs have an intrinsic affinity towards the free interface (e.g.,~if they are hydrophobic or amphiphilic) or if a droplet has a sufficiently large number of INPs. Indeed, variations in INP concentration among different droplets have already been shown to result in large uncertainties in rate estimates~\cite{KnopfClimAtmosSci2020}. Our findings suggest that INP-free surface proximity can result in even larger uncertainties, and quantifying its likelihood is critical to obtaining more reliable heterogeneous nucleation rate estimates. The same framework can be used to probe nucleation in other liquids suspected of surface freezing, such as silicon~\cite{LiNatMater2009}. 

Finally, the theoretical approach proposed in this work can be applied to other scenarios in which crystalline nuclei might simultaneously form on multiple interfaces-- or interfacial patches-- with different chemistries or topographies. This could, for instance, occur in 'Janus` slit pores comprised of different types of confining surfaces\cite{KumarLangmuir2018}. A more interesting scenario, however, emerges when a single interface is comprised of multiple distinct "patches``, or nanoscale regions with differing chemistries and ice nucleating propensities. Such patchy surfaces can emerge in a wide variety of  systems, such as complex organic aerosols\cite{KnopfACSEarthSpaceChem2018}, block oligomers\cite{BiggsNatComm2017} and polymers\cite{JellinekColloidPolymSci1978} and ice nucleating~\cite{HudaitJACS2019} and antifreeze~\cite{HajiAkbariPNAS2016} proteins. Such patchy surfaces and coatings have garnered increased interest recently due to advancements in various top-down techniques\cite{SeoJNanosciNanotech2014} and bottom-up approaches\cite{KimChemRev2010}, such as block copolymer self-assembly\cite{RuizScience2008, LuoMacromolecules2012} which have made their precision fabrication possible.  In principle, the applicability of the theoretical framework proposed in this work to study nucleation on such patchy surfaces does not depend on their particular chemistries and mechanical and topographical properties. However, such patches sometimes resemble free interfaces,} such as hydrophobic patches on a protein~\cite{PandeySciAdv2016}, henceforth making the interfacial proximity of the type discussed here even more {\color{black}salient. Our findings  demonstrate that different aspects of confinement can be harnessed to realize nonclassical nucleation behavior.

\subsection*{Supporting Information (SI)}

\noindent
Further computational details, including the derivation of geometric features of the utilized CNT-based theory,  the approach for estimating model parameters and uncertainty analysis, implementation details of jFFS, and further information about system setup are all included in the Supplementary Information.

\section*{Acknowledgements}
\noindent 
 We thank P. G. Debenedetti, T. Koop, V. Molinero, F. Caupin, D. Knopf, A. Michaelides and G. Sosso for useful discussions. These calculations were performed on the Yale Center for Research Computing. This work used the Extreme Science and Engineering Discovery Environment (XSEDE), which is supported by National Science Foundation grant no. ACI-1548562~\cite{TownsCompSciEng2014}.
A.H.-A. gratefully acknowledges the support of the National Science Foundation CAREER Award (Grant No. CBET-1751971).

\bibliographystyle{apsrev}

\bibliography{References}

\clearpage

\appendix 

\section*{SUPPLEMENTARY INFORMATION}

\setcounter{figure}{0}
\setcounter{table}{0}
\renewcommand{\thefigure}{S\arabic{figure}}
\renewcommand{\thetable}{S\arabic{table}}

\section{CNT-Based Theoretical Description of Nucleation in Thin Films}
\label{SI-section:CNT}

\subsection{Estimating Volumes and Liquid-Exposed Surface Areas}
\label{SI-subsection:volumes}

\noindent
Here, we present a detailed geometric analysis of how  to calculate the volume, $V_{\text{hg}}$, and liquid-exposed surface area, $S_{\text{hg}}$, of hourglass-shaped nuclei of Figure~\ref{Figure 3}b. 
In doing so, we need to separately consider two distinct scenarios depending on the geometrical feasibility of an explicit liquid-exposed cylindrical bridge. In principle, such a bridge will be possible only if  the two spherical caps do not intersect at all (Figure~\ref{SI-fig:schematic-hourglass}a) or that their plane of intersection lies in between the centers of the two spheres (Figure~\ref{SI-fig:schematic-hourglass}b). Therefore, no such bridge will be feasible for a nucleus like the one depicted in Figure~\ref{SI-fig:schematic-hourglass}c. For nuclei of the type depicted in Figure~\ref{SI-fig:schematic-hourglass}a-b, $V_{\text{hg}}$ and $S_{\text{hg}}$ can be expressed as follows,
\begin{eqnarray}\label{funnel_volume}
V_{\text{hg}}^{\theta_w,\theta_f}(r_w,r_f, r_c, l) &=& \sum_{i\in\{w,f\}} \left[
V_{\text{cap}}^{\theta_i}(r_i) - V_{\text{subcap}}(r_i, r_c)
\right]\notag\\
&& +\pi r_c^2L^{\theta_i,\theta_f}(r_w,r_f,r_c,l) 
\label{eq:case-1-V} \\
S_{\text{hg}}^{\theta_w,\theta_f}(r_w,r_f, r_c, l) &=& \sum_{i\in\{w,f\}} \left[
S_{\text{cap}}^{\theta_i}(r_i) - S_{\text{subcap}}(r_i, r_c)
\right]\notag\\
&& +2\pi r_cL^{\theta_i,\theta_f}(r_w,r_f,r_c,l)
\label{eq:case-1-S}
\end{eqnarray}
Here, $V_{\text{cap}}^{\theta_i}(r_i)$ and $S_{\text{cap}}^{\theta_i}(r_i)$ are the volume and surface area of the spherical cap forming at interface $i\in\{w,f\}$. $V_{\text{subcap}}(r_i, r_c)$ and $S_{\text{subcap}}(r_i, r_c)$, however, correspond to the volume and surface area of the "subcap`` region at the intersection of the spherical cap with the cylinder. These quantities can be estimated from,
\begin{eqnarray}
V_{\text{cap}}^{\theta_i}(r_i) &=& \frac43\pi r_i^3f_c(\theta_i)
\label{eq:V-cap}\\
S_{\text{cap}}^{\theta_i}(r_i) &=& 2\pi r_i^2(1-\cos\theta_i)
\label{eq:S-cap}\\
V_{\text{subcap}}(r_i, r_c) &=& \frac\pi3 H^2(r_i,r_c)\left[3r_i-H(r_i,r_c)\right] 
\label{eq:V-subcap}\\
S_{\text{subcap}}(r_i,r_c) &=& 2\pi r_iH(r_i,r_c)
\label{eq:S-subcap}
\end{eqnarray}
with $f_c(\theta_i)$, the potency factor, and $H(r_i,r_c)$, the height of the subcap region given by,
\begin{eqnarray}
f_c(\theta_i) &=& \tfrac14(1-\cos\theta_i)^2(2+\cos\theta_i)
\label{eq:potency}\\
H(r_i,r_c)&=&r_i-\sqrt{r_i^2-r_c^2} 
\label{eq:subcap-height}
\end{eqnarray}
Finally $L^{\theta_w,\theta_f}(r_w,r_f,r_c,l)$ is the height of the cylindrical bridge and can be estimated from, 
	\begin{widetext}
	\begin{eqnarray}
	L^{\theta_w,\theta_f}(r_w,r_f,r_c,l) &=& \left\{
	\begin{array}{ll}
	l + \sum_{i\in\{w,f\}}\left[H(r_i,r_c)-r_i(1-\cos\theta_i)\right] &  \mathcal{Q}^{\theta_w, \theta_f}(r_w,r_f,l) \ge 0\\
	d^{\theta_w, \theta_f}(r_w, r_f, l) - \sum_{i\in\{w,f\}}d_{\text{cyl},i} & \mathcal{Q}^{\theta_w,\theta_f} (r_w,r_f,l) < 0 \\
	\end{array}
	\right.
	\label{eq:cylinder-length}
	\end{eqnarray}
	\end{widetext}
with $\mathcal{Q}^{\theta_w, \theta_f}(r_w,r_f,l)$ given by,
\begin{eqnarray}
\mathcal{Q}^{\theta_w, \theta_f}(r_w,r_f,l)&=& l - \sum_{i\in\{w,f\}}r_i(1 - \cos\theta_i),
\label{eq:sphere-dist}
\end{eqnarray}
Note that $\mathcal{Q}^{\theta_w, \theta_f}(r_w,r_f,l)$ has different geometrical interpretations for non-intersecting and intersecting caps. In the case of non-intersecting caps, $\mathcal{Q}^{\theta_w, \theta_f}(r_w,r_f,l)(>0)$  corresponds to the minimum distance between the caps. For intersecting caps, however, $\mathcal{Q}^{\theta_w, \theta_f}(r_w,r_f,l)(<0)$ is a measure of the extent of penetration of the caps. $d^{\theta_w, \theta_f}(r_w, r_f, l)$ is the relative elevation of the center of the free interfacial cap with respect to that of the wall-based cap and is given by,
\begin{eqnarray}
d^{\theta_w,\theta_f}(r_w,r_f,l) &=& l + \sum_{i\in\{w,f\}}r_i\cos\theta_i,
\label{eq:sphere-center-dist}
\end{eqnarray}
and $d_{\text{cyl},i}=\sqrt{r_i^2-r_c^2}$ is the distance between the center of sphere $i$ and the closer base of the cylinder. 

Note that the validity of Eqs.~(\ref{eq:case-1-V}), (\ref{eq:case-1-S}), (\ref{eq:V-subcap}), (\ref{eq:S-subcap}), (\ref{eq:subcap-height}) and (\ref{eq:cylinder-length}) is predicated upon the geometrical feasibility of a liquid-exposed cylinder. This further depends on the relative elevation of plane of intersection of two caps with respect to center of the first sphere, which is given by,
\begin{eqnarray}
z_{\text{int}}^{\theta_w,\theta_f}(r_w,r_f,l) = \frac{\left[d^{\theta_w,\theta_f}(r_w,r_f,l)\right]^2-r_f^2+r_w^2}{2d^{\theta_w,\theta_f}(r_w,r_f,l)} \notag\\ 
\end{eqnarray}
Note that a cylinder will only be feasible if $\mathcal{Q}\ge0$ or $|z_{\text{int}}|<|d|$. Otherwise, $V_{\text{hg}}$ and $S_{\text{hg}}$ can be estimated from,
\begin{eqnarray}\label{net_volume}
V_{\text{hg}}^{\theta_w,\theta_f}(r_w, r_f, l) &=&  \sum_{i\in\{w,f\}}V_{\text{cap}}^{\theta_i}(r_i) - V_{\text{lens}}^{\theta_w,\theta_f} (r_w, r_f, l) \notag\\
&&
 \\
S_{\text{hg}}^{\theta_w,\theta_f}(r_w, r_f, l) &=& \sum_{i\in\{w,f\}}\Big[S_{\text{cap}}^{\theta_i}(r_i) -2\pi r_ih_i^{\theta_w,\theta_f}(r_w, r_f, l)\Big] \notag\\
&& 
\end{eqnarray}
Here, $V_{\text{lens}}^{\theta_w,\theta_f} (r_w, r_f, l)$ is the volume of the intersection of the two caps, and is given by,
\begin{widetext}
\begin{eqnarray}
V_{\text{lens}}^{\theta_w,\theta_f} (r_w, r_f, l) &=&  
\frac{\pi\left[\sum_{i\in\{w,f\}}r_i-d^{\theta_w,\theta_f}(r_w,r_f,l)\right]^2}{12d^{\theta_w,\theta_f}(r_w,r_f,l)}\notag\\
&& \times\Big\{\left[d^{\theta_w,\theta_f}(r_w,r_f,l)\right]^2-3\left(r_w-r_f\right)^2 +2d^{\theta_w,\theta_f}(r_w,r_f,l)(r_w+r_f)\Big\}
\end{eqnarray} 
\end{widetext}
$h_i^{\theta_w,\theta_f}(r_w, r_f, l)'s$, however, are the heights of the smaller spherical caps that constitute the lens, and are given by,
\begin{eqnarray}
h_w^{\theta_w,\theta_f}(r_w, r_f, l) &=& \frac{r_f - r_w + d^{\theta_w,\theta_f}(r_w,r_f,l)}{2d^{\theta_w,\theta_f}(r_w,r_f,l)}\notag\\
&& \times \left[r_w + r_f - d^{\theta_w,\theta_f}(r_w,r_f,l)\right]  \notag\\
&& \\
h_f^{\theta_w,\theta_f}(r_w, r_f, l) &=& \frac{r_w - r_f + d^{\theta_w,\theta_f}(r_w,r_f,l)}{2d^{\theta_w,\theta_f}(r_w,r_f,l)}\notag\\
&& \times\left[r_w + r_f - d^{\theta_w,\theta_f}(r_w,r_f,l)\right] \notag\\
&& 
\end{eqnarray}

\subsection{Constraints on $\textit{r}_\textit{w}, \textit{r}_\textit{f}$ and $\textit{r}_\textit{c}$}
\label{SI-subsection:constraints}

\noindent
As mentioned in the main text, a spherical cap can only intersect with the opposing surface within the base of intersection of the other cap. Likewise, a cylindrical bridge cannot touch the two surfaces outside the bases of intersections of the two spherical caps. These requirements, which assure a constant contact angle at each interface, further limit the permissible values of $r_w, r_f$ and $r_c$. More particularly, $\rho^{\theta_i}(r_i,l)$, the radius of the circle forming at an opposite interface, is given by,
\begin{eqnarray}
\rho^{\theta_i} (r_i,l) &=& \left\{
\begin{array}{ll}
\sqrt{r_i^2-(r_i\cos\theta_i+l)^2} & r_i(1-\cos\theta_i) \ge l\\
0 & r_i(1-\cos\theta_i) < l
\end{array}
\right. \notag\\
&& 
\end{eqnarray}
It can be easily shown that $r_w,r_f$ and $r_c$ need to satisfy the following inequalities:
\begin{subequations}
\begin{eqnarray}
\rho^{\theta_w}(r_w,l) &<& r_f\sin\theta_f \label{eq:rho-w} \\
\rho^{\theta_f}(r_f,l) &<& r_w\sin\theta_w \label{eq:rho-f} 
\end{eqnarray}
\begin{eqnarray}
r_c &\le& \left\{
\begin{array}{ll}
\min\{r_w,r_f\}& \theta_w,\theta_f>\frac\pi2\\
\min\{r_w\sin\theta_w,r_f\sin\theta_f\} & \theta_w, \theta_f\le\frac\pi2\\
\min\{r_w,r_f\sin\theta_f\} & \theta_w>\frac\pi2, \theta_f\le\frac\pi2\\
\min\{r_w\sin\theta_w,r_f\} & \theta_w\le\frac\pi2, \theta_f>\frac\pi2
\end{array}
\right.\notag\\
&& 
\label{eq:ineq-rc}
\end{eqnarray}
\end{subequations}
Note that a 'virtual` cylinder (i.e.,~a cylinder with $L=0$) might exist in accordance with Eq.~(\ref{eq:ineq-rc}). 

\vspace{30pt}

\subsection{Constraints on Maximum Volume for Obtuse Contact Angles}
\label{SI-subsection:obtuse}

\noindent
In this section, we prove the assertion that we mentioned in the main text, that if both contact angles are obtuse, $V_{\text{hg}}$ is bounded from above. We start with rewriting (\ref{eq:rho-w}) and (\ref{eq:rho-f}) as,
\begin{eqnarray}
\sin^2\theta_w-2\alpha\omega\cos\theta_w-\alpha^2\omega^2 &<& \alpha^2\sin^2\theta_f\label{eq:ineq-1}\\
\alpha^2\sin^2\theta_f-2\alpha^2\omega\cos\theta_f-\alpha^2\omega^2 &<& \sin^2\theta_w\label{eq:ineq-2}
\end{eqnarray}
whereby $\alpha:=r_f/r_w$ and $\omega:=l/r_f$. Eliminating $\sin^2\theta_w$ from (\ref{eq:ineq-1}) and (\ref{eq:ineq-2}) yields:
\begin{eqnarray}
&&-2\alpha^2\omega\cos\theta_f-\alpha^2\omega^2<2\alpha\omega\cos\theta_w+\alpha^2\omega^2\notag\\
&& \implies -\alpha\cos\theta_f-\cos\theta_w<\alpha\omega\notag\\
&& \implies \boxed{
-r_f\cos\theta_f-r_w\cos\theta_w<l
}\label{eq:ineq-final}
\end{eqnarray}
Note that if $\theta_f$ and $\theta_w$ are obtuse, the lefthand side of~(\ref{eq:ineq-final}) will be positive, and a strictly increasing function of $r_f$ and $r_w$. This will imply that for sufficiently large $r_f$'s and $r_w$'s, this inequality will be violated, and thus the size of the nucleus will be bounded from above.

\subsection{Numerical Estimation of $\Delta{\textit{G}}^{\textit{l}}_{\text{hg}}(\textit{N})$ and $\Delta{\textit{G}}^{\text{*}}$} \label{SI-subsection:CNT:numerics}

\noindent
In order to estimate the nucleation barrier $\Delta{G}^*$, we first solve the optimization problem given in Eq.~(\ref{eq:GvsN}) by calculating $\Delta{G}_{\text{hg}}^l(r_w,r_f,r_c)$ on a three-dimensional grid in the $(r_w,r_f,r_c)$ space with a uniform grid resolution of $0.0025$~nm in each dimension. For the 1.2 nm- and 1.8 nm-thick films, the grid is comprised of $1000^3$ points, while a slightly larger grid is used for thicker films ($1200^3$ and $1600^3$ for 2.4-nm and 3.6-nm films, respectively). These grid sizes are chosen so that the nuclei corresponding to $\Delta{G}^*$ fall within the grid for each film. At each grid point, $N$, the number of molecules within the nucleus is estimated as $N(r_w,r_f,r_c):=n_{\text{int}} [\rho_sV_{\text{hg}}^l(r_w,r_f,r_c)]$ where $n_{\text{int}}[x]$  is the closest integer to $x\in\mathbb{R}^{\ge0}$. $\Delta{G}^l_{\text{hg}}(N)$ is estimated by minimizing $\Delta{G}_{\text{hg}}^l(r_w,r_f,r_c)$ over all the grid points with the same $N$. $\Delta{G}^*$ is then estimated as the local maximum in $\Delta{G}^l_{\text{hg}}(N)$. {\color{black}In our calculations, we use $l$ values that are determined by identifying the largest $z$ values in the  density profiles of Figure~\ref{SI-fig:density} at which the local density drops to 50\% of the bulk density. These 'exact` thicknesses do not deviate from the approximate values referred in the paper by more than 10\%. The results presented in Figures~3c and 6b, in particular, are obtained with $l=1.24$~nm and 1.29~nm, respectively.  We, however, prefer to refer to the approximate values in the text in order to make comparisons between different INPs and system types more straightforward. Conducting our CNT-based theoretical calculation using the  approximate $l$'s does not change the model predictions significantly. }

\section{Parameter Estimation and Uncertainty Analysis}\label{SI-sec:param-analysis}

\subsection{Contact Angles\label{SI-subsection:contact-angles}} 

\noindent As mentioned in the main text, classical nucleation theory assumes that crystalline nuclei maintain a fixed contact angle at the nucleating surface. Assuming that crystalline nuclei are spheres in the case of homogeneous nucleation and spherical caps in the case of heterogeneous nucleation, the size of the respective critical nuclei will be given by:
\begin{eqnarray}
N_{\text{hom}}^* &=& \frac{32\pi\gamma_{sl}^3}{3\rho_s^3|\Delta\mu|^3}\label{eq:N-homo}\\
N_{\text{het}}^* &=& N_{\text{hom}}^*f_c(\theta_c)\label{eq:N-het}
\end{eqnarray}
Here, $\gamma_{sl}, \rho_s$ and $|\Delta\mu|$ are the liquid-solid surface tension, solid number density and the chemical potential difference between the solid and the liquid, respectively, and $f_c(\theta_c)$ is the potency factor {\color{black} given by Eq.~(\ref{eq:potency})}. In order to estimate the contact angles, we first compute critical nucleus sizes for homogeneous and heterogeneous nucleation, and then use Eq.~(\ref{eq:N-het}) to estimate the potency factor and the contact angle. This approach assumes the validity of CNT for heterogeneous nucleation, which has been validated in earlier studies of heterogeneous ice nucleation on graphene surfaces.\cite{CabrioluPRE2015} We compute critical nucleus sizes from committor probabilities, which can be estimated from FFS transition probabilities as follows,
\begin{eqnarray}
p_c({\color{black}\xi}_j) &=& \prod_{i=j}^{N-1} P({\color{black}\xi}_{i+1}|{\color{black}\xi}_i).
\label{eq:p_c}
\end{eqnarray}
The critical nucleus size $N^*$ is determined from the following fit,
\begin{eqnarray}
p_c({\color{black}\xi})&=&\frac12\Bigg\{1+\text{erf}~[a({\color{black}\xi}-N^*)]\Bigg\},\label{eq:p-c-fit}
\end{eqnarray}
which assures that $p_c(N^*)=0.5$. This analysis is predicated upon the assumption that the FFS order parameter is  a good reaction coordinate for nucleation. As discussed elsewhere,\cite{LupiNature2017} any reasonable measure of the size of the largest crystalline nucleus (including the one utilized here) is indeed a good reaction coordinate for nucleation. Therefore, our proposed approach will provide a reasonable estimate of $N^*$.   {\color{black}All the reported error bars in $N^*$ correspond to the 95\% confidence intervals of the parameter $N^*$ estimated from the fit given in Eq.~(\ref{eq:p-c-fit}).} 

For homogeneous nucleation in the bulk SW21 system, we use the rate calculation at $T_0=174$~K, reported in Ref.~\citenum{GianettiPCCP2016}. Using Eq.~(\ref{eq:p-c-fit}), we estimate $N_{\text{hom}}^*(174~\text{K})$ to be  {\color{black}$631.4\pm0.3$}. We assume within our theory that nucleation at the free interface is heterogeneous in nature, even though in reality it is more likely pseudo-heterogeneous.\cite{ReissJPhysChemA2002, HajiAkbariPNAS2017} We therefore determine the potency factor from $f_c(\theta_f)=N_{\text{film}}^*/N_{\text{bulk}}^*$ wherein $N_{\text{film}}^*$ is the critical nucleus size for nucleation in a 5-nm-thick freestanding thin film of the SW21 liquid. Using the rate calculations at Ref.~\citenum{GianettiPCCP2016} and Eq.~(\ref{eq:p-c-fit}), we estimate $N_{\text{film}}^*={\color{black}598.1\pm0.8}$ at 174~K, which yields a contact angle of $\theta_f={\color{black}136.2^{\circ}\pm0.3}^{\circ}$.

In order to compute $\theta_w$, we use our rate calculations in 3.6-nm thick films as a reference for unperturbed heterogeneous nucleation. This choice is warranted since a well-developed bulk-like region separates the free interface and the INP in 3.6-nm films, and the observed nucleation mechanism does not deviate from what is expected for classical heterogeneous nucleation. Since all our heterogeneous nucleation rate calculations are conducted at temperatures other than $T_0=174~\text{K}$, Eq.~(\ref{eq:N-het}) cannot be utilized without estimating $N^*_{\text{hom}}$ at the respective temperature. Assuming that $\gamma_{sl}$ and $\rho_s$ are not strong functions of temperature, $N_{\text{hom}}^*(T)$ can be estimated from Eq.~(\ref{eq:N-homo}),
\begin{eqnarray}
N_{\text{hom}}^*(T) &=& N_{\text{hom}}^*(T_0)\frac{|\Delta\mu(T_0)|^3}{|\Delta\mu(T)|^3}.
\end{eqnarray} 
Note that $\Delta\mu(T)$ can be accurately computed using thermodynamic integration, 
\begin{eqnarray}
\Delta\mu (T) &=& T\int_T^{T_m} \frac{h_l(\overline{T})-h_s(\overline{T})}{\overline{T}^2}d\overline{T},
\end{eqnarray}
where $h_s$ and $h_l$ are molar enthalpies of the (hexagonal) crystal and the supercooled liquid computed from $NpT$ MD simulations at 1~atm, and $T_m=206$~K is the melting temperature for the SW21 model. Figure~\ref{SI-fig:dmuvsT} depicts $\Delta\mu(T)$ for the SW21.   Using the corrective scheme outlined above, we estimate $N_{\text{hom}}^*$ to be $44{\color{black}8.1}\pm{\color{black}0.2}$ and $308{\color{black}.2}\pm{\color{black}0.2}$ at 170~K and 165~K, respectively. Similarly, we estimate $N_{\text{het}}^*$ to be {\color{black}$216.2\pm1.8$} and ${\color{black}251.7\pm1.2}$ for graphene wall (at 170~K) and structureless LJ 9-3 wall (at 165~K), respectively. This corresponds to $\theta_w={\color{black}88.6^{\circ}\pm0.3}^{\circ}$ for graphene and $\theta_w={\color{black}117.0^{\circ}\pm0.4}^{\circ}$ for the structureless LJ 9-3 wall.

{\color{black}

\subsection{Predictions of the CNT-based Theory} \label{SI-subsection:CNT-based-error-bars} 

\noindent
The analysis provided here is based on the assumption that the main source of  uncertainty in the predictions of our CNT-based model is the uncertainty in the underlying model parameters. In principle, the input parameters that are most prone to uncertainty are $\gamma_{sl}$, $\theta_w$ and $\theta_f$. For $\gamma_{sl}$, we use the 95\% confidence interval reported in Ref.~\citenum{GianettiPCCP2016}, while the error bars in $\theta_w$ and $\theta_f$ are estimated in Section~\ref{SI-subsection:contact-angles} using CNT. All these error bars are, however, estimated assuming the validity of CNT, and the high quality of the classification criterion employed for detecting solid-like molecules. This analysis therefore does not take into account such difficult-to-quantify uncertainties. 

With these error bars at hand, we use sensitivity analysis to determine how such uncertainties translate into uncertainties in model predictions. More precisely, we compute the respective quantity of interest at the center and the vertices of a hyper-cube that spans the confidence intervals for the relevant input parameters. We employ this approach due to lack of analytical solutions for the optimization problems posed in the main text, including in Eq.~(\ref{eq:GvsN}).  We are, in particular, interested in determining the uncertainties in the following model predictions: 
\begin{enumerate}
\item $\Delta{G}_{\text{diff}}$ at the original-- unperturbed-- $\theta_f$ and the associated $N^*$ value.
\item The perturbed $\theta_f$ needed for reproducing the observed enhancement in nucleation.
\item $N^*$ estimated at the corrected $\theta_f$.
\end{enumerate}
We compute $\Delta{G}_{\text{diff}}$ at the original-- unperturbed-- $\theta_f$ at $(\gamma_{sl}\pm\delta\gamma_{sl}, \theta_w\pm\delta\theta_w,\theta_f\pm\delta\theta_f)$. In order to estimate the error bars in the perturbed $\theta_f$ and the associated $N^*$, we compute $\Delta{G}_{\text{diff}}$ for a large number of $\theta_f$'s but at $(\gamma_{sl},\theta_w), (\gamma_{sl},\theta_w\pm\delta\theta_w), (\gamma_{sl}\pm\delta\gamma_{sl},\theta_w)$ and  $(\gamma_{sl}\pm\delta\gamma_{sl},\theta_w\pm\delta\theta_w)$. This gives us nine functions that are depicted in Figures 3c and 6b in the main text. The bounds of the confidence intervals in the corrected $\theta_f$'s and $N^*$'s are thus determined as the minimum and maximum values obtained from crossing these nine functions with the target $\Delta{G}_{\text{diff}}$. The findings of this analysis are presented in Table~\ref{tab:theory-errorbars}. 

}

{\color{black}
\subsection{Planar RDFs, and $q_\text{3}$ and Nearest Neighbor Count Distributions}\label{SI-subsection:estimate:RDFq3}

\noindent
The uncertainties in planar RDFs, and $q_3$ and nearest neighbor count distributions are estimated as follows. First, at each state point, $N_t$ independent MD simulations of the supercooled liquid are conducted, with each simulation initiated from a configuration that has been quenched from a different high-temperature configuration and conducted for 0.5~ns. A per-trajectory profile $f_i(\eta)$ is computed for each trajectory $1\le i\le N_t$ where $\eta$ is the coordinate of interest, i.e.,~$r$, $q_3$ and $N_{\text{ngb}}$ for planar RDFs, and $q_3$ and nearest neighbor count distributions, respectively. $f_i(\eta)$'s are computed by binning the  liquid along the $z$ direction, and averaging the profiles obtained for the bins that are within the target density peak, i.e.,~those whose local densities deviate from the peak density by less than 10\%. The average profile $\overline{f}(\eta)$ is then computed as,
\begin{eqnarray}
\overline{f}(\eta) &=& \frac{1}{N_t}\sum_{i=1}^{N_t} f_i(\eta), \label{eq:f-eta-avg}
\end{eqnarray}
The error bar in $\overline{f}(\eta)$ is then estimated as,
\begin{eqnarray}
\delta\overline{f}^2 (\eta) &=& \frac{1}{N_t-1}\sum_{i=1}^{N_t}\left[f_i(\eta)-\overline{f}(\eta)\right]^2, \label{eq:f-eta-error}
\end{eqnarray}
The error bars reported in Figures~4, 5 and \ref{SI-fig:struct-intermediate} are all 95\% confidence intervals, i.e.,~$2\sqrt{\delta\overline{f}^2(\eta)}$, and are obtained from a minimum of 100 independent trajectories. For the ultrathin SW21 films wherein nucleation is fast enough to occur frequently during unbiased MD trajectories, the order parameter $\xi(\cdot)$ is  computed at the end of each trajectory, and the trajectories with a $\xi_\text{final}>25$ are not included in Eqs.~(\ref{eq:f-eta-avg}) and (\ref{eq:f-eta-error}). 
}

\section{Technical Details of jFFS Calculations}
\label{SI-section:jFFS}

\noindent The values of ${\color{black}\xi}_A$ and ${\color{black}\xi}_\text{0}$ are determined using the approach described in Ref.~\citenum{HajiAkbariFilmMolinero2014}.
The successive interfaces in jFFS are determined according to the scheme of Ref.~\citenum{HajiAkbariJCP2018}, wherein each target FFS-interface is chosen to lie beyond the maximum ${\color{black}\xi}$ value reached in the previous iteration. For each rate calculation, we explore the supercooled liquid basin for a minimum of 0.4~$\mu$s. We terminate each FFS iteration after a minimum of 2,000 crossings, while we require more crossings for earlier iterations (a minimum of 4,000 and 3,000 at the first and the second interface, respectively). Error bars are computed using the approach described in Ref.~\citenum{AllenFrenkel2006}. Due to the interfacial nature of heterogeneous nucleation, we report all rates in $\text{m}^{-2}\cdot\text{s}^{-1}$, unlike several earlier works that have reported them in $\text{m}^{-3}\cdot\text{s}^{-1}$. All system sizes and computed rates are summarized in Table~\ref{tab:system-summary}. 

{\color{black}
\section{Derivation of Eq.~(\ref{eq:Jv-uniform}) in the Main Text}
\label{SI-section:derivation}

\noindent
Assuming that INPs are uniformly distributed with a droplet of radius $r_0$ and $\mathcal{J}_a$ is independent of $\Phi$ and is given by,
\begin{eqnarray}
\mathcal{J}_a &=& \left\{
\begin{array}{ll}
\mathcal{J}_a^{(s)} &~~~ s\le s_0\\
\mathcal{J}_a^{(i)} &~~~ s>s_0
\end{array}
\right.
\end{eqnarray}
Eq.~(\ref{eq:Jv-general}) can be expressed as,
\begin{eqnarray}
\mathcal{J}_v(r_0) &=& \frac{3an}{r_0^3}\left\{\frac13(r_0-s_0)^3\mathcal{J}_a^{(i)}
+\frac{1}{3}\left[r_0^3-(r_0-s_0)^3
\right]\mathcal{J}_a^{(s)}
\right\}\notag\\
&=& an\left\{
\left(1-\frac{s_0}{r_0}
\right)^3\mathcal{J}_a^{(i)} +
\left[
1 - \left(1-\frac{s_0}{r_0}
\right)^3
\right]\mathcal{J}_a^{(s)}
\right\}\notag\\
&\overset{\text{(a)}}{\approx} & an\mathcal{J}_a^{(i)} \left[
1 + 3\frac{s_0}{r_0}\frac{\mathcal{J}_a^{(s)}}{\mathcal{J}_a^{(i)}}
\right] = an\mathcal{J}_a^{(i)} (1+3\alpha\varsigma)\notag
\end{eqnarray}
Note that (a) follows from using the Taylor expansion $(1-x)^n=1-nx+O(x^2)$ for $|x|\ll1$ and neglecting higher order terms. \textcolor{black}{We denote the enhancement factor corresponding to a uniform distribution of  $\mathcal{J}_a(s)$ within $s\le s_0$ by $\chi_u$.}
}

{\color{black}

Now consider the case in which $\mathcal{J}_a(s)$ depends on the number of liquid layers that separate the surface of the INP from the free interface. Let $q_m$ be the thickness of an $m$-layer liquid film separating a flat INP from the free interface. It is reasonable to expect that $0=q_0<q_1<\cdots<q_k=s_0$ where $s_0$ is the largest $s$ for which $\mathcal{J}_a(s)$ differs considerably from the immersion rate $\mathcal{J}_a^{(i)}$. Also, physical intuition will imply that $q_{m}-q_{m-1}$ will not be a strong function of $m$ as the thickness of each liquid layer will be mostly constant. Assuming that $\mathcal{J}_a(s) = \mathcal{J}_{a,m}^{(s)}$ for $q_{m-1}<s\le q_{m}$, the  apparent volumetric rate within a droplet of radius $r_0\gg s_0$ will be given by,
\begin{eqnarray}
\mathcal{J}_v(r_0) &=& an\mathcal{J}_{a}^{(i)}\left[
1 + 3\sum_{m=1}^k\frac{q_m-q_{m-1}}{r_0} \frac{\mathcal{J}_{a,m}^{(s)}}{\mathcal{J}_{a}^{(i)}}
\right] \notag\\
&=& an\mathcal{J}_{a}^{(i)}\left[1+3\sum_{m=1}^k \alpha_m\varsigma_m
\right]
\end{eqnarray}
with $\alpha_m:=(q_m-q_{m-1})/r_0$ and $\varsigma_m:=\mathcal{J}_{a,m}^{(s)}/\mathcal{J}_{a}^{(i)}$. The enhancement factor $\chi$ will thus be given by,
\begin{eqnarray}
\chi&=&3\sum_{m=1}^k \alpha_m\varsigma_m.
\end{eqnarray}
In general, $\chi$  will be dominated by the term corresponding to the largest $\varsigma_m$. Therefore even if $\mathcal{J}_{a,m}^{(s)}$ is a strong function of $m$, $\chi$ will still be orders of magnitude larger than unity as long as some of its constituent terms are orders of magnitude larger than unity.

Now suppose a "worst-case scenario`` in which $\varsigma_k \gg \varsigma_{k-1} \approx \cdots \approx \varsigma_1\sim1$,~i.e.,~that nucleation is only enhanced within a supported film comprised of exactly $k$ liquid layers, and not for thicker or thinner films.  Under such a scenario, the enhancement factor will be given by,
\begin{eqnarray}
\chi &=& 3\sum_{m=1}^k \alpha_m\varsigma_m \approx  3\alpha_k\varsigma_k = 3\alpha_k\frac{q_k-q_{k-1}}{s_0}\frac{s_0}{r_0}\label{eq:modified-enhancement-factor}
\end{eqnarray}
This implies that $\chi$ will only differ from $\chi_u$ by an algebraic factor $(q_k-q_{k-1})/s_0\sim 1/k$ while its order of magnitude will remain unchanged. Therefore, the kinetic enhancement during inside-out contact freezing will still be astronomical even if the effect observed in this work is only limited to 1.2-nm films, and not for films that are thicker or thinner. 
}

\clearpage

\begin{table*}
	\centering
	\caption{\label{tab:system-summary}{Summary of system sizes and nucleation rates for the supported films considered in this work. $\textit{S}$ and $\textit{l}$ refer to the surface area of the INP, and the thickness of the supported film, respectively. $\textit{N}_{\textit{p}}$ and $\textit{N}_\textit{g}$, however, are the number of mW/SW21 molecules and graphene atoms, respectively. Note that structureless INPs, by definition, have no explicit atoms. }}
	\begin{tabular}{ccrccrrr}
	\hline\hline
	{\color{black}System} &
	{\color{black}INP} &
	{\color{black}$\textit{S}~(\text{nm}^{\text{2}})$} & 
	{\color{black}$\textit{l}$~(nm)} & 
	{\color{black}$\textit{T}$~(K)} & 
	{\color{black}$\textit{N}_{\text{p}}~~$} & 
	{\color{black}$\textit{N}_{\text{g}}~~$} & 
	{\color{black}$\text{log}_{\text{10}}\textit{R}~(\text{m}^{-\text{2}}\cdot\text{s}^{-\text{1}})$} \\
	\hline
	{\color{black} mW} &
	{\color{black} Graphene} & 	
	{\color{black} 40.8623} & 
	{\color{black} 1.2} & 
	{\color{black} 235} & 
	{\color{black} 1,600} & \
	\textcolor{black}{1,560}  & 
	{\color{black} +20.5397$\pm$0.2224} \\
	{\color{black} mW} &
	{\color{black} Graphene} & 	
	{\color{black} 40.8623} & 
	{\color{black} 1.8} & 
	{\color{black} 235} & 
	{\color{black} 2,400} & 
	\textcolor{black}{1,560}& 
	{\color{black} +20.1649$\pm$0.4596}\\	
	{\color{black} mW} &
	{\color{black} Graphene} & 	
	{\color{black} 40.8623} & 
	{\color{black} 2.4} & 
	{\color{black} 235} & 
	{\color{black} 3,200} & 
	\textcolor{black}{1,560}&  
	{\color{black} +21.7089$\pm$0.3647}\\
	{\color{black} mW} &
	{\color{black} Graphene} & 	
	{\color{black}40.8623} & 
	{\color{black}3.6} & 
	{\color{black}235}  & 
	{\color{black}4,800} & 
	\textcolor{black}{1,560}&  
	{\color{black}+21.3811$\pm$0.3664} \\
	\hline
	{\color{black} mW} &
	{\color{black} Graphene} & 	
	{\color{black}40.8623} & 
	{\color{black}1.2} & 
	{\color{black}240} & 
	{\color{black}1,600} & 
	\textcolor{black}{1,560}& 
	{\color{black}+17.3360$\pm$0.1593} \\
	{\color{black} mW} &
	{\color{black} Graphene} & 	
	{\color{black}40.8623} & 
	{\color{black}1.8} & 
	{\color{black}240} & 
	{\color{black}2,400} & 
	\textcolor{black}{1,560}& 
	{\color{black}+17.3122$\pm$0.1709}\\
	{\color{black} mW} &
	{\color{black} Graphene} & 	
	{\color{black}40.8623} & 
	{\color{black}2.4} & 
	{\color{black}240} & 
	{\color{black}3,600} & 
	\textcolor{black}{1,560}& 
	{\color{black}+17.1180$\pm$0.1376} \\
	{\color{black} mW} &
	{\color{black} Graphene} & 	
	{\color{black}40.8623} & 
	{\color{black}3.6} & 
	{\color{black}240} & 
	{\color{black}4,800} & \textcolor{black}{1,560} & 
	{\color{black}+18.4664$\pm$0.2055}\\
	\hline
	{\color{black} SW21} &
	{\color{black} Graphene} & 	
	{\color{black}40.8623} & 
	{\color{black}1.2} & 
	{\color{black}170}  & 
	{\color{black}~1,600}  & 
	\textcolor{black}{1,560} & 
	{\color{black}+24.9746$\pm$}{\color{black}0.0507} \\
	SW21\footnote{This calculation has been conducted using conventional MD and the mean first passage time (MFPT) method of Ref.~\citenum{WedekindJChemPhys2007}.}
	{\color{black} Graphene} & 	
	{\color{black}163.4498} & 
	{\color{black}1.2} & 
	{\color{black}170}  & 
	{\color{black}~6,400}  & 
	\textcolor{black}{6,240} & 
	{\color{black}+25.3349$\pm$}{\color{black}0.0740} \\
	{\color{black} SW21} &
	{\color{black} Graphene} & 	
	{\color{black} 163.4498} & 
	{\color{black} 1.8} & 
	{\color{black} 170}  &
	{\color{black} ~9,600} & 
	\textcolor{black}{6,240} & 
	{\color{black} +19.6054$\pm$0.1208}\\
	{\color{black} SW21} &
	{\color{black} Graphene} & 	
	{\color{black} 163.4498} & 
	{\color{black}2.4} & 
	{\color{black}170} & 
	{\color{black}12,800} & 
	\textcolor{black}{6,240} & 
	{\color{black}+19.3989$\pm$0.0786}\\
	{\color{black} SW21} &
	{\color{black} Graphene} & 	
	{\color{black}163.4498} & 
	{\color{black}3.6} & 
	{\color{black}170}  & 
	{\color{black}19,200} & 
	\textcolor{black}{6,240} & 
	{\color{black}+18.6513$\pm$}{\color{black}0.1304} \\
	\hline
	{\color{black} mW} &
	{\color{black} LJ 9-3} & 	
	{\color{black}40.3332} & 
	{\color{black}1.2} & 
	{\color{black}220} & 
	{\color{black}1,600} & 
	{\color{black}N/A} & 
	{\color{black}+20.0712$\pm$}{\color{black}0.0804} \\
	{\color{black} mW} &
	{\color{black} LJ 9-3} &
	{\color{black}40.3332} & 
	{\color{black}3.6} & 
	{\color{black}220} & 
	{\color{black}4,800} & 
	{\color{black}N/A} & 
	{\color{black}+20.8655$\pm$}{\color{black}0.0748} \\
	\hline
	{\color{black} SW21} &
	{\color{black} LJ 9-3} & 
	{\color{black}40.3332} & 
	{\color{black}1.2} & 
	{\color{black}165} & 
	{\color{black}1,600}  & 
	{\color{black}N/A} & 
	{\color{black}+22.9647$\pm$}{\color{black}0.2608} \\
	{\color{black} SW21} &
	{\color{black} LJ 9-3} & 
	{\color{black}162.5089} & 
	{\color{black}3.6} & 
	{\color{black}165}  & 
	{\color{black}19,200} & 
	{\color{black}N/A} & 
	{\color{black}+6.8265$\pm$}{\color{black}0.4484} \\
	\hline\hline
	\end{tabular}
\end{table*}

\begin{figure*}
\centering
\includegraphics[width=.8767\textwidth]{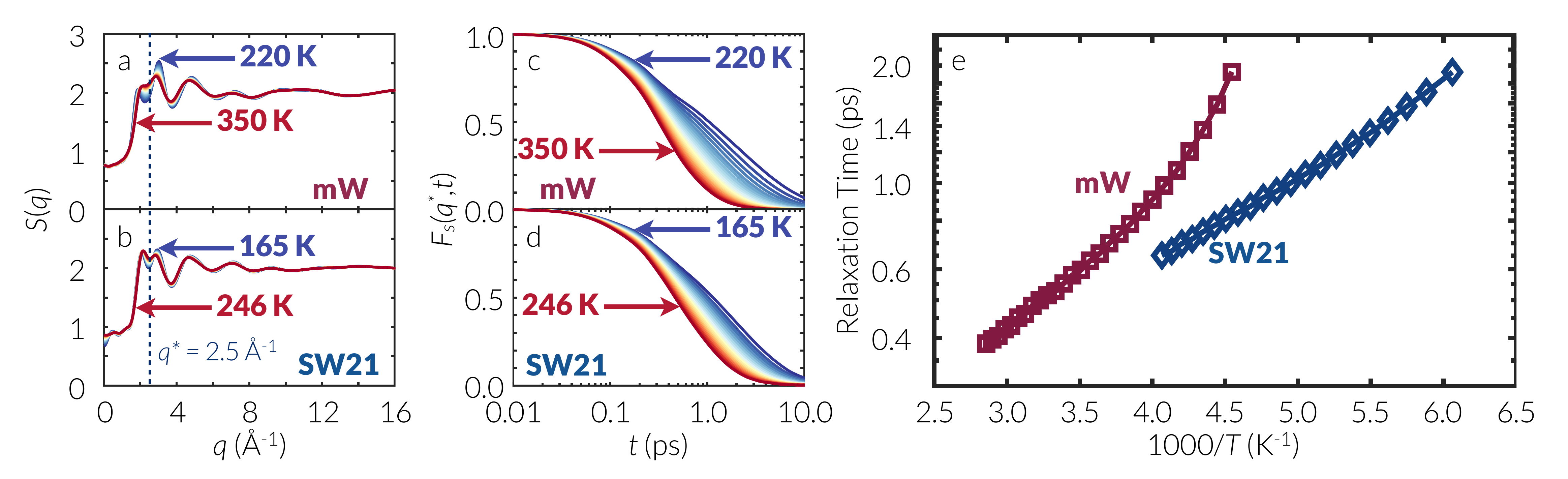} 
\caption{\color{black}(a-b) Structure factor, $S(q)$, and (c-d) self-intermediate scattering function, $F_S(q^*,t)$, for (a,c) mW and (b,d) SW21 liquids at $p=0$~bar. $q^*=2.5~\AA^{-1}$~for both (c) and (d) is close to the first valley of $S(q)$. Each simulation is conducted for a minimum of 9.8~ns within a cubic simulation box comprised of 1728 mW or SW21 molecules. (e) Relaxation times computed by fitting a stretched exponential to the $\beta$-relaxation parts of $F_S(q^*,t)$ in both (c,d). \label{fig:relaxation-times}}
\end{figure*}

\begin{figure*}
\centering
\includegraphics[width=\textwidth]{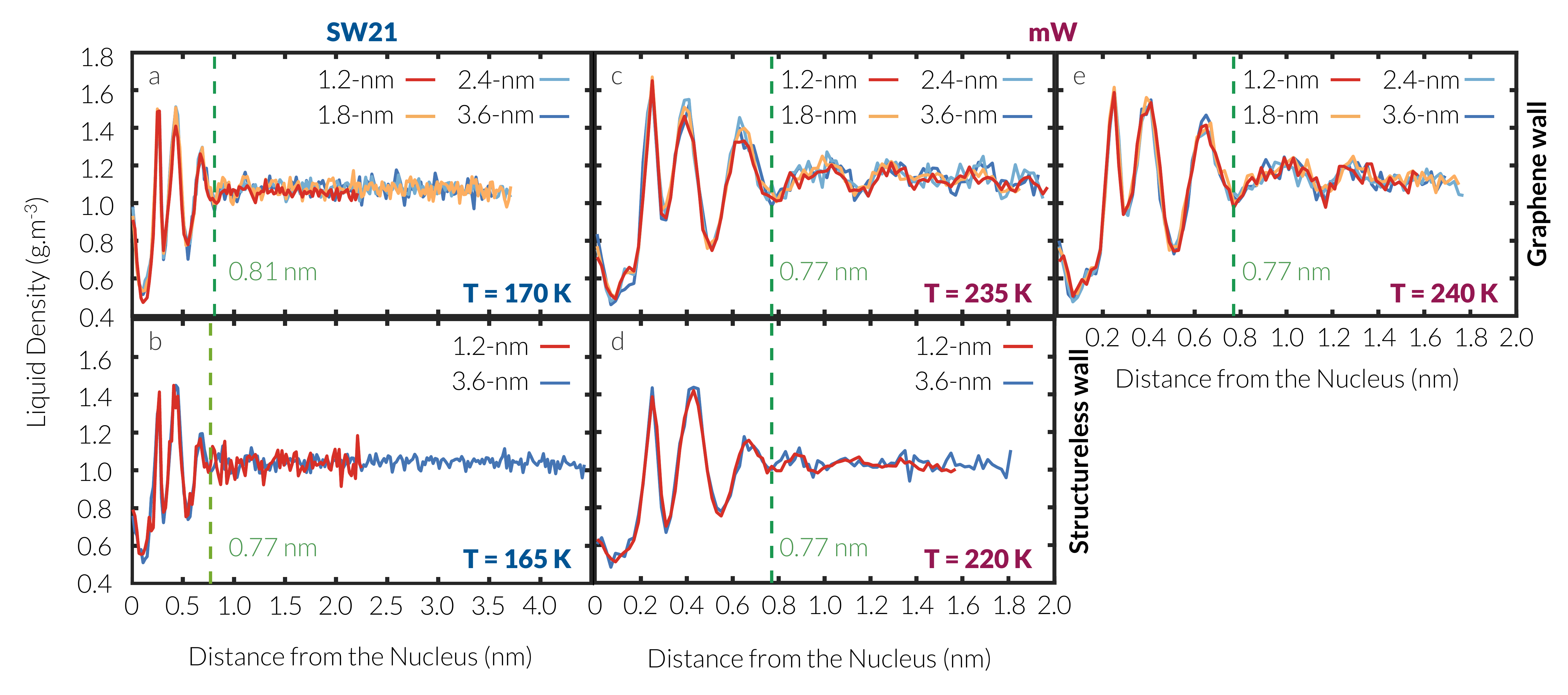} 
\caption{\color{black}The inter-image liquid density as a function of the distance from the critical nuclei for all rate calculations conducted in this work. Consistent with Ref.~\citenum{HussainJCP2020p}, $\textbf{u}$, the shortest vector that connects the molecules within a critical nucleus to those in its closest periodic image is identified, and the average density along $\textbf{u}$ is computed by enumerating the number of  molecules located within the intersection of a cylinder along $\textbf{u}$ and the INP. The cylinder has a radius of 0.32~nm and 0.345~nm for mW and SW21, respectively. The first three peaks correspond to the diffusely structured liquid around the nuclei, while the existence of the central plateau region corresponds to lack of appreciable finite size effects. The inter-image plateau densities given in Table~\ref{tab:finite-size} are computed for the plateau regions in (a-e).  \label{SI-fig:finite-size-effects} }
\end{figure*}

\begin{table*}
	{
	\centering
	\caption{\label{tab:finite-size}{\color{black}Inter-image plateau densities computed using the approach described in Ref.~\citenum{HussainJCP2020p}. Each reported plateau density $\rho_{\textit{p}}^\textit{n}$ is computed for the critical configurations in each respective system with inter-image connections completely located within the $\textit{n}$th liquid layer. $\rho_{\textit{l}}^\textit{n}$, the average supercooled liquid density within the $\textit{n}$th  layer, is computed using the density profiles depicted in Figure~\ref{SI-fig:density}. All error bars correspond to 95\% confidence intervals with the number of independent samples equal to the number of distinct ancestors at $\xi_\text{0}$. The N/A entries correspond to situations in which either no inter-image connection exists within a particular layer, or if all configurations with such connections share the same ancestor at $\xi_\text{0}$ resulting in an undetermined confidence interval. Note that the differences between plateau densities and liquid densities are very small, corresponding to the existence of weak finite size effects. }}
	\begin{tabular}{ccccccccc}
	\hline\hline
	\color{black} System & 
	\color{black} \textit{T}~[K] & 
	\color{black} $\textit{l}$~[nm]& 
	\color{black} $\rho_{\textit{p}}^\text{1}~[\text{g}\cdot\text{cm}^{\text{-3}}]$& 
	\color{black} $\rho_{\textit{l}}^\text{1}~[\text{g}\cdot\text{cm}^{\text{-3}}]$& 
	\color{black} $\rho_{\textit{p}}^\text{2}~[\text{g}\cdot\text{cm}^{\text{-3}}]$& 
	\color{black} $\rho_{\textit{l}}^\text{2}~[\text{g}\cdot\text{cm}^{\text{-3}}]$& 
	\color{black} $\rho_{\textit{p}}^\text{3}~[\text{g}\cdot\text{cm}^{\text{-3}}]$& 
	\color{black} $\rho_{\textit{l}}^\text{3}~[\text{g}\cdot\text{cm}^{\text{-3}}]$ \\
	\hline
	\color{black} SW21/Graphene & 
	\color{black} 170 & 
	\color{black} 1.2 & 
	\color{black} 1.0716$\pm$0.0026 & 
	\color{black} 1.0624$\pm$0.0007 &
	\color{black} 1.0715$\pm$0.0481 & 
	\color{black} 1.0998$\pm$0.0007 &
	\color{black} 1.0659$\pm$0.0088 & 
	\color{black} 1.0538$\pm$0.0014 \\
	\color{black} SW21/Graphene & 
	\color{black} 170 & 
	\color{black} 1.8 & 
	\color{black} 1.0665$\pm$0.0015 & 
	\color{black} 1.0632$\pm$0.0001 &
	\color{black} N/A & 
	\color{black} 1.2326$\pm$0.0002 &
	\color{black} N/A & 
	\color{black} 1.0114$\pm$0.0001 \\
	\color{black} SW21/Graphene & 
	\color{black} 170 & 
	\color{black} 2.4 & 
	\color{black} 1.0656$\pm$0.0014 & 
	\color{black} 1.0638$\pm$0.0002 &
	\color{black}  N/A & 
	\color{black} 1.2482$\pm$0.0002 &
	\color{black} N/A & 
	\color{black} 1.0129$\pm$0.0001 \\
	\color{black} SW21/Graphene & 
	\color{black} 170 & 
	\color{black} 3.6 & 
	\color{black} 1.0664$\pm$0.0017 & 
	\color{black} 1.0635$\pm$0.0002 &
	\color{black}  N/A & 
	\color{black} 1.2471$\pm$0.0002 &
	\color{black} N/A & 
	\color{black} 1.0099$\pm$0.0001 \\
	\hline
	\color{black} mW/Graphene & 
	\color{black} 235 & 
	\color{black} 1.2 & 
	\color{black} 1.3549$\pm$0.0017 & 
	\color{black} 1.3522$\pm$0.0002 &
	\color{black} 1.0155$\pm$0.0073 & 
	\color{black} 1.0154$\pm$0.0002 &
	\color{black} 1.0393$\pm$0.0112 & 
	\color{black} 1.0109$\pm$0.0027 \\
	\color{black} mW/Graphene & 
	\color{black} 235 & 
	\color{black} 1.8 & 
	\color{black} 1.3788$\pm$0.0021 & 
	\color{black} 1.3754$\pm$0.0002 &
	\color{black} 1.0077$\pm$0.0023 & 
	\color{black} 1.0020$\pm$0.0001 &
	\color{black} N/A & 
	\color{black} 1.0302$\pm$0.0002 \\
	\color{black} mW/Graphene & 
	\color{black} 235 & 
	\color{black} 2.4 & 
	\color{black} 1.3516$\pm$0.0066 & 
	\color{black} 1.3521$\pm$0.0002 &
	\color{black} 1.0157$\pm$0.0051 & 
	\color{black} 1.0154$\pm$0.0002 &
	\color{black} 1.0315$\pm$0.0702 & 
	\color{black} 1.0240$\pm$0.0003 \\
	\color{black} mW/Graphene & 
	\color{black} 235 & 
	\color{black} 3.6 & 
	\color{black} 1.3562$\pm$0.0018 & 
	\color{black} 1.3521$\pm$0.0002 &
	\color{black} 1.0279$\pm$0.0017 & 
	\color{black} 1.0154$\pm$0.0002 &
	\color{black} N/A & 
	\color{black} 1.0242$\pm$0.0004 \\
	\hline
	\color{black} mW/Graphene & 
	\color{black} 240 & 
	\color{black} 1.2 & 
	\color{black} 1.4097$\pm$0.0063 & 
	\color{black} 1.4082$\pm$0.0015 &
	\color{black} 1.0157$\pm$0.0103& 
	\color{black} 1.0106$\pm$0.0020 &
	\color{black} 1.0226$\pm$0.0627 & 
	\color{black} 1.0256$\pm$0.0057 \\
	\color{black} mW/Graphene & 
	\color{black} 240 & 
	\color{black} 1.8 & 
	\color{black} 1.3840$\pm$0.0028 & 
	\color{black} 1.3817$\pm$0.0012 &
	\color{black} 1.0186$\pm$0.0059& 
	\color{black} 1.0204$\pm$0.0009 &
	\color{black} 1.0129$\pm$0.0410 & 
	\color{black} 1.0280$\pm$0.0029 \\
	\color{black} mW/Graphene & 
	\color{black} 240 & 
	\color{black} 2.4 & 
	\color{black} 1.4353$\pm$0.0027 & 
	\color{black} 1.4312$\pm$0.0011 &
	\color{black} 1.0153$\pm$0.0060& 
	\color{black} 1.0054$\pm$0.0010 &
	\color{black} N/A & 
	\color{black} 1.0285$\pm$0.0023 \\
	\color{black} mW/Graphene & 
	\color{black} 240 & 
	\color{black} 3.6 & 
	\color{black} 1.4033$\pm$0.0009 & 
	\color{black} 1.3972$\pm$0.0023 &
	\color{black} 1.0203$\pm$0.0013& 
	\color{black} 1.0146$\pm$0.0012 &
	\color{black} N/A & 
	\color{black} 1.0200$\pm$0.0017 \\
	\hline
	\color{black} SW21/LJ 9-3 & 
	\color{black} 165 & 
	\color{black} 1.2 & 
	\color{black} 1.0589$\pm$0.0166 & 
	\color{black} 1.0489$\pm$0.0003 &
	\color{black} N/A & 
	\color{black} 1.1017$\pm$0.0003 &
	\color{black} 1.0932$\pm$0.0320 & 
	\color{black} 1.0617$\pm$0.0023 \\
	\color{black} SW21/LJ 9-3 & 
	\color{black} 165 & 
	\color{black} 3.6 & 
	\color{black} 1.0476$\pm$0.0008 & 
	\color{black} 1.0444$\pm$0.0001 &
	\color{black} N/A & 
	\color{black} 1.1027$\pm$0.0001 &
	\color{black} N/A & 
	\color{black} 1.0572$\pm$0.0001 \\
	\hline
	\color{black} mW/LJ 9-3 & 
	\color{black} 220 & 
	\color{black} 1.2 & 
	\color{black} 1.1413$\pm$0.0043 & 
	\color{black} 1.1222$\pm$0.0002 &
	\color{black} 1.0585$\pm$0.0175 & 
	\color{black} 1.0404$\pm$0.0002 &
	\color{black} 1.0216$\pm$0.0323 & 
	\color{black} 1.0063$\pm$0.0004 \\
	\color{black} mW/LJ 9-3 & 
	\color{black} 220 & 
	\color{black} 3.6 & 
	\color{black} 1.1422$\pm$0.0044 & 
	\color{black} 1.1256$\pm$0.0003 &
	\color{black} 1.0501$\pm$0.0163 & 
	\color{black} 1.0384$\pm$0.0003 &
	\color{black} 1.0002$\pm$0.0230 & 
	\color{black} 0.9950$\pm$0.0003 \\
	\hline\hline
	\end{tabular}
	}
\end{table*}

\begin{figure*}
\centering
\includegraphics[width=.7878\textwidth]{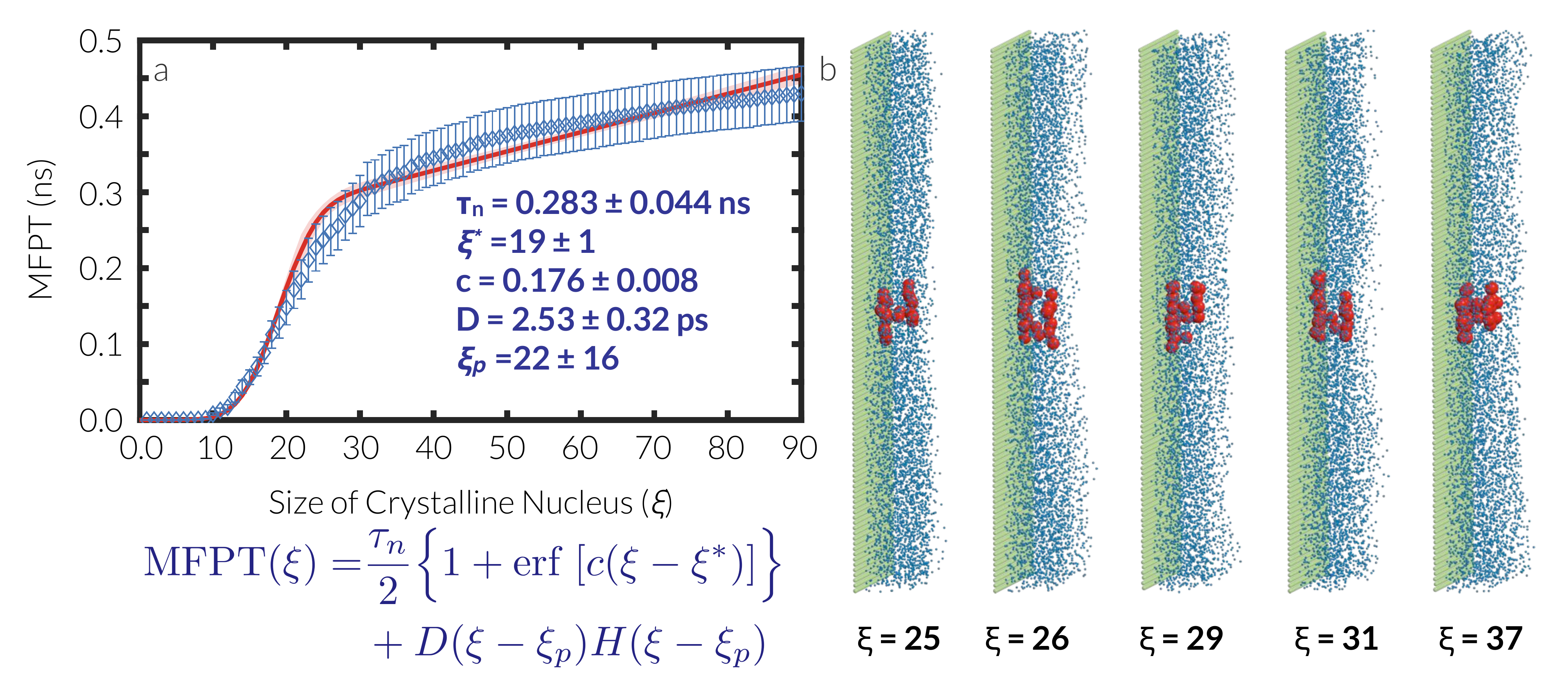} 
\caption{\color{black} (a) Mean first passage time (MFPT) for nucleation within a 1.2-nm thick SW21 film in the vicinity of a larger graphene INP (i.e.,~one with $S=163.4498~\text{nm}^2$) obtained from 188 independent trajectories with a combined duration of 165~ns at 170~K. The term $D(\xi-\xi_p)H(\xi-\xi_p)$ corresponds to crystal growth with $H(\xi)$ the Heaviside function. The nucleation rate is estimated from $\mathcal{J}=1/(\tau_nS)$ and is given by $\log_{10}\mathcal{J}~[\text{m}^{-2}\text{s}^{-1}]=25.3349\pm0.0740$. (b) Several representative configurations along the nucleation pathway, all adopting an hourglass geometry.  \label{SI-fig:larger-INP} }
\end{figure*}

\begin{table*}
	{
	\centering
	\caption{\label{tab:homogeneous-rates}{\color{black}Comparison of heterogeneous nucleation rates computed in this work to the respective homogeneous nucleation rates at the same temperature. In order to make such a comparison possible, we divide the areal heterogeneous rates by the film thickness to convert them to volumetric rates. For the mW system, the homogeneous rates are based on those computed in Ref.~\citenum{HajiAkbariJCP2018} or their CNT extrapolations. In the SW21 system, however, all the estimates are based on the CNT extrapolation given in Eq.~(16) and Fig.~5 of Ref.~\citenum{GianettiPCCP2016}.}}
	\begin{tabular}{cccccc} 
	\hline\hline
	{\color{black}System/INP Type} &
	{\color{black} \textit{T}~[K]} &
	{\color{black} Film Thickness [nm]} &
	{\color{black} $\text{log}_{\text{10}}\textit{R}_\text{het,s}~[\text{m}^{-\text{2}}\cdot\text{s}^{-\text{1}}]$} &
	{\color{black} $\text{log}_{\text{10}}\textit{R}_\text{het,v}~[\text{m}^{-\text{3}}\cdot\text{s}^{-\text{1}}]$} &
	{\color{black} $\text{log}_{\text{10}}\textit{R}_\text{hom}~[\text{m}^{-\text{3}}\cdot\text{s}^{-\text{1}}]$} \\
	\hline
	{\color{black}mW/Graphene} &
	{\color{black}235} &
	{\color{black}1.2} &
	{\color{black}20.5397$\pm$0.2224} &
	{\color{black}29.4605$\pm$0.2224} &
	\multirow{4}{*}{\color{black}5.6294$\pm$3.5901}\\
	{\color{black}mW/Graphene} &
	{\color{black}235} &
	{\color{black}1.8} &
	{\color{black}20.1649$\pm$0.4596} &
	{\color{black}28.9096$\pm$0.4596} &
	\\
	{\color{black}mW/Graphene} &
	{\color{black}235} &
	{\color{black}2.4} &
	{\color{black}21.7089$\pm$0.3647} &
	{\color{black}30.3287$\pm$0.3647} &
	\\
	{\color{black}mW/Graphene} &
	{\color{black}235} &
        {\color{black}3.6} &
	{\color{black}21.3811$\pm$0.3664} &
	{\color{black}29.8248$\pm$0.3664} &
	\\
	\hline
	{\color{black}mW/Graphene} &
	{\color{black}240} &
	{\color{black}1.2} &
	{\color{black}17.3360$\pm$0.1593} &
	{\color{black}26.2568$\pm$0.1593} &
	\multirow{4}{*}{\color{black}-7.1521$\pm$5.8757}\\
	{\color{black}mW/Graphene} &
	{\color{black}240} &
	{\color{black}1.8} &
	{\color{black}17.3122$\pm$0.1709} &
	{\color{black}26.0569$\pm$0.1709} &
	\\
	{\color{black}mW/Graphene} &
	{\color{black}240} &
	{\color{black}2.4} &
	{\color{black}17.1180$\pm$0.1376} &
	{\color{black}25.7378$\pm$0.1376} &
	\\
	{\color{black}mW/Graphene} &
	{\color{black}240} &
	{\color{black}3.6} &
	{\color{black}18.4664$\pm$0.2055} &
	{\color{black}26.9101$\pm$0.2055} &
	\\
	\hline
	{\color{black}SW21/Graphene} &
	{\color{black}170} &
	{\color{black}1.2} &
	{\color{black}24.9746$\pm$0.0507} &
	{\color{black}33.8954$\pm$0.0507} &
	\multirow{4}{*}{\color{black}-18.1079$\pm$6.2371}\\
	{\color{black}SW21/Graphene} &
	{\color{black}170} &
	{\color{black}1.8} &
	{\color{black}19.6054$\pm$0.1208} &
	{\color{black}28.3501$\pm$0.1208} &
	\\
	{\color{black}SW21/Graphene} &
	{\color{black}170} &
	{\color{black}2.4} &
	{\color{black}19.3989$\pm$0.0786} &
	{\color{black}28.0187$\pm$0.0786} &
	\\
	{\color{black}SW21/Graphene} &
	{\color{black}170} &
	{\color{black}3.6} &
	{\color{black}18.6513$\pm$0.1304} &
	{\color{black}27.0950$\pm$0.1304} &
	\\
	\hline
	{\color{black}mW/LJ 9-3} &
	{\color{black}220} &
	{\color{black}1.2} &
	{\color{black}20.0712$\pm$}{\color{black}0.0804} &
	{\color{black}28.9920$\pm$}{\color{black}0.0804} &
	\multirow{2}{*}{\color{black}24.8672$\pm$0.2454}\\
	{\color{black}mW/LJ 9-3} &
	{\color{black}220} &
	{\color{black}3.6} &
	{\color{black}20.8655$\pm$}{\color{black}0.0748} &
	{\color{black}29.3092$\pm$}{\color{black}0.0748} &
	\\
	\hline
	{\color{black}SW21/LJ 9-3} &
	{\color{black}165} &
	{\color{black}1.2} &
	{\color{black}22.9647$\pm$}{\color{black}0.2608} &
	{\color{black}31.8855$\pm$}{\color{black}0.2608} &
	\multirow{2}{*}{\color{black}-6.9572$\pm$4.8333}\\
	{\color{black}SW21/LJ 9-3} &
	{\color{black}165} &
	{\color{black}3.6} &
	{\color{black}6.8265$\pm$}{\color{black}0.4484} &
	{\color{black}15.2701$\pm$}{\color{black}0.4484} &
	\\
	\hline
	\hline
	\end{tabular}
	}
\end{table*}

\begin{figure*}
\centering
\includegraphics[width=.7467\textwidth]{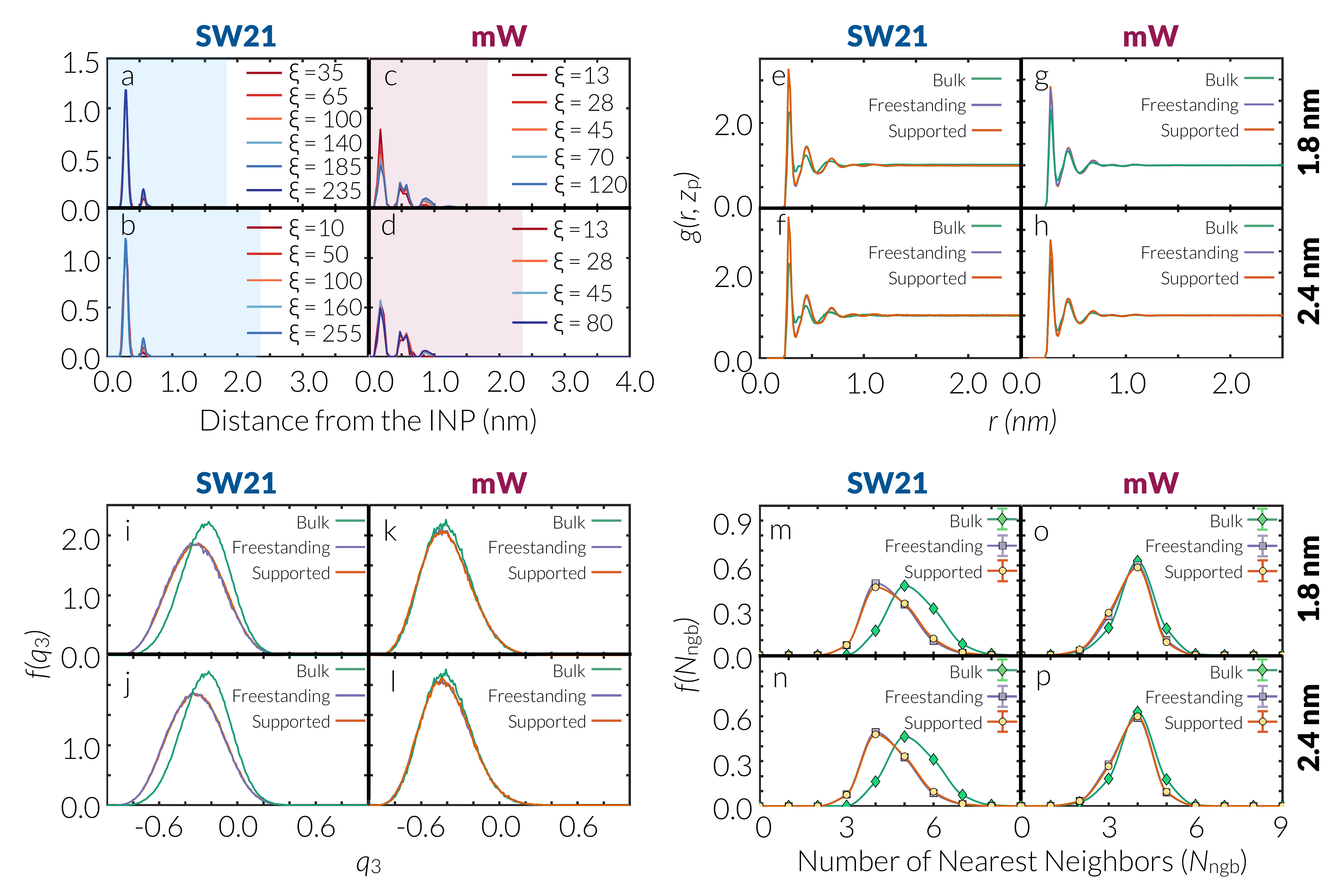} 
\caption{Nucleation mechanism and the characterization of free interfacial regions for intermediate-thickness graphene-supported SW21 and mW films at 170~K and 235~K, respectively. (a-d) Histograms of the $\textit{z}$ coordinates of the molecules belonging to the largest crystalline nucleus in surviving configurations of (a-b) SW21 and (c-d) mW films, all pointing to a mechanism consistent with classical heterogeneous nucleation observed in 3.6-nm films. Shaded regions correspond to the geometric spreads of the supported films.   (e-h ) Planar RDF's, and (i-l) $\textit{q}_{\textit{3}}$ and (m-p) nearest neighbor count distributions for free interfaces of supported SW21 (e-f, i-j, m-n) and mW (g-h, k-l, o-p) films. Note that there is no distinction between free interfacial properties of supported and freestanding films. {\color{black}Error bars in (e-h) and (m-p) are thinner than the curves and smaller  than the symbols, respectively. The areas under the curves in (i-l) are normalized to unity, and the shades correspond to error bars.}}
\label{SI-fig:struct-intermediate}
\end{figure*}

\begin{figure*}
	\centering
	\includegraphics[width=.7832\textwidth]{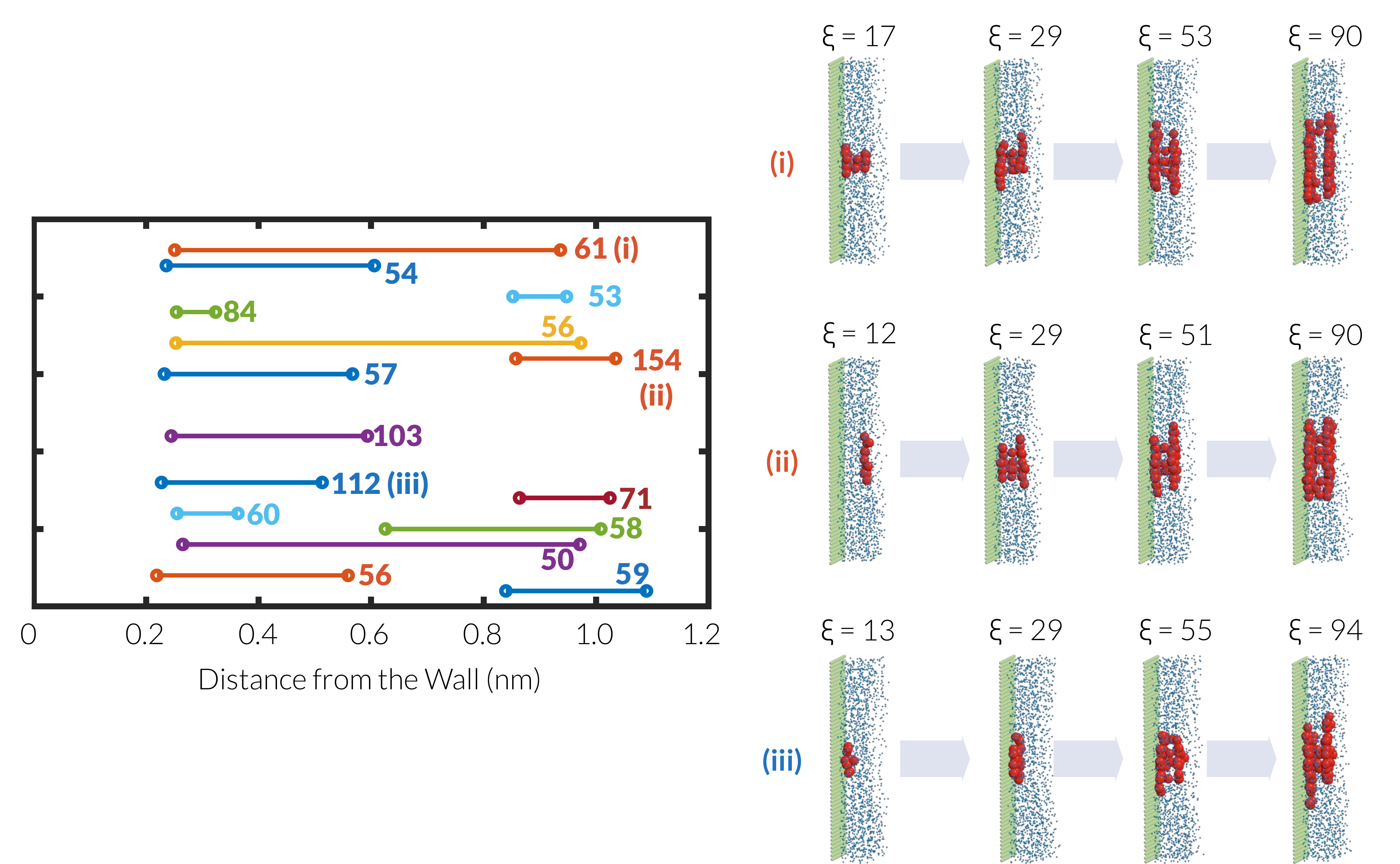}
	\caption{Geometric spread of the largest crystalline nucleus for the most prolific surviving configurations at ${\color{black}\xi}_{\text{0}}$=12 in the 1.2-nm thick SW21 graphene-supported film. Each label corresponds to the number of progeny of the configuration at the last milestone. While several of these configurations have nuclei that span the entire film. many of them emerge at either interface and subsequently grow to form hourglass-shaped nuclei. Three representative pathways are depicted on the right for nuclei that start (i) at both interface, (ii) at the free interface, and (iii) at the INP.}
	\label{SI-fig:geom-spread}
\end{figure*}

\begin{figure*}
	\centering
	\includegraphics[width=.7679\textwidth]{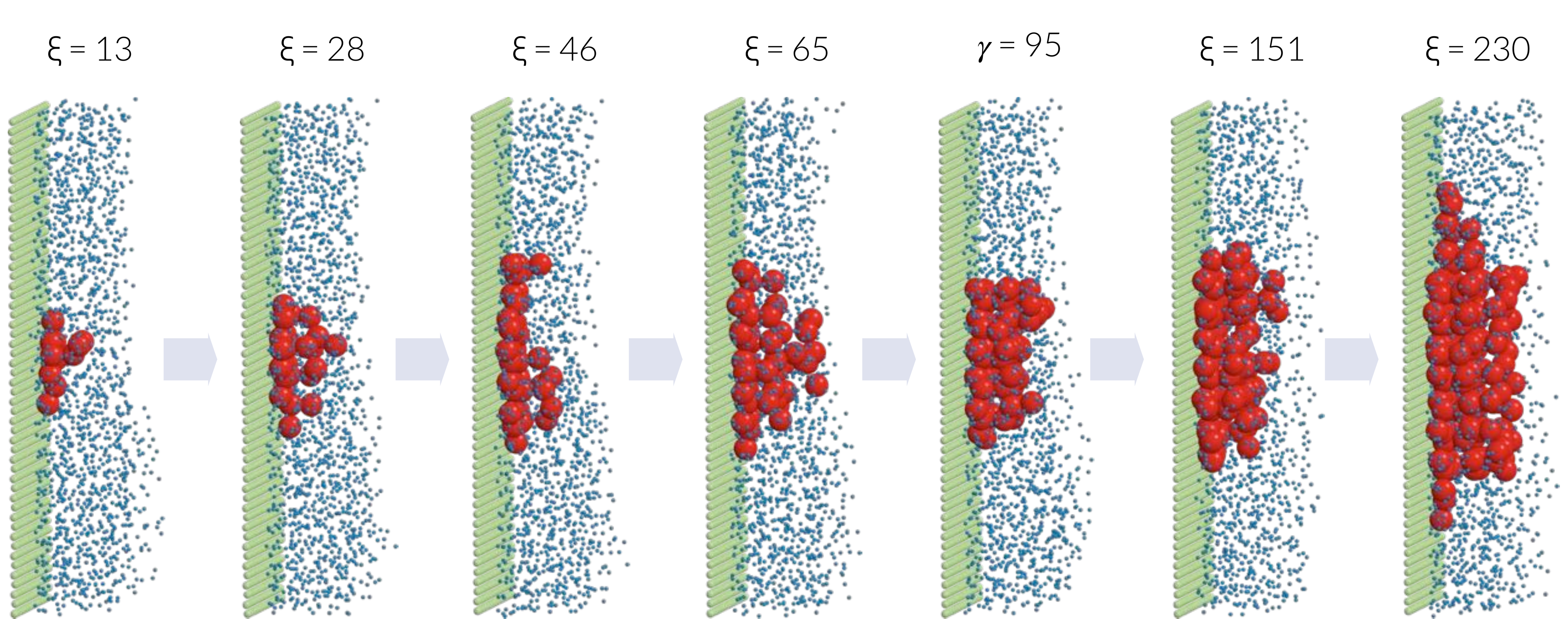}
	\caption{A typical nucleation pathway in an ultrathin mW film at 235~K. Each configuration is a progeny of the configuration to its left.}
	\label{SI-fig:pathway-mW}
\end{figure*}

\begin{figure*}
\centering
\includegraphics[width=.5478\textwidth]{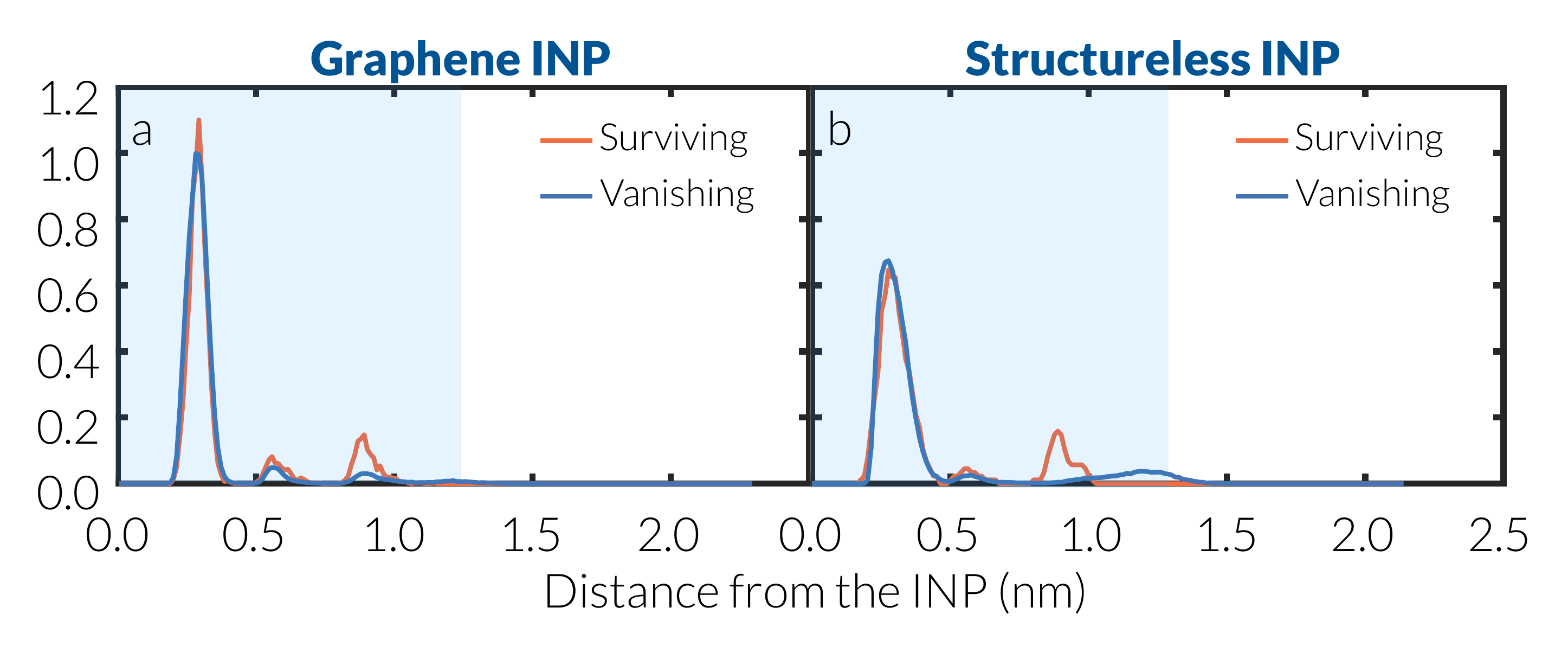} 
\caption{\color{black}Histograms of the $\textit{z}$ coordinates of the molecules belonging to the largest crystalline nucleus in surviving and vanishing configurations at $\xi_0$ within an ultrathin SW21 film in the vicinity of (a) graphene and (b) the structureless INP.  Shaded regions correspond to the geometric spreads of the supported films.   The areas under the curves are normalized to unity.\label{SI-fig:surviving-vanishing}}
\end{figure*} 

\begin{figure*}
\centering
\includegraphics[width=.2991\textwidth]{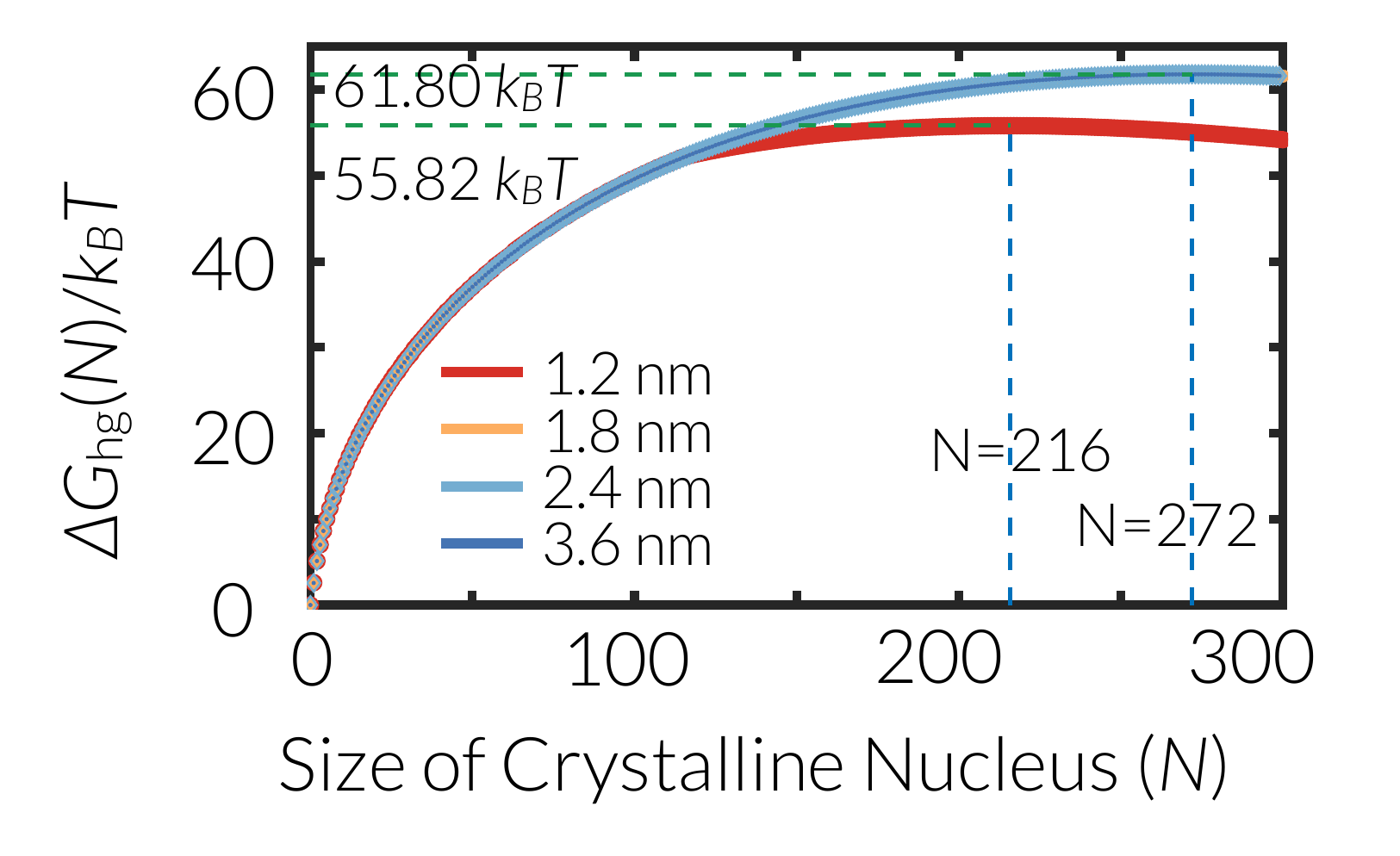} 
\caption{\color{black} $\Delta\textit{G}_{\text{hg}}(\textit{N})$ for SW21 films of different thicknesses predicted from the CNT-based theory discussed in the main text.\label{SI-fig:DeltaG-CNT}}
\end{figure*} 

 \begin{figure*}
\centering
\includegraphics[width=0.4469\textwidth]{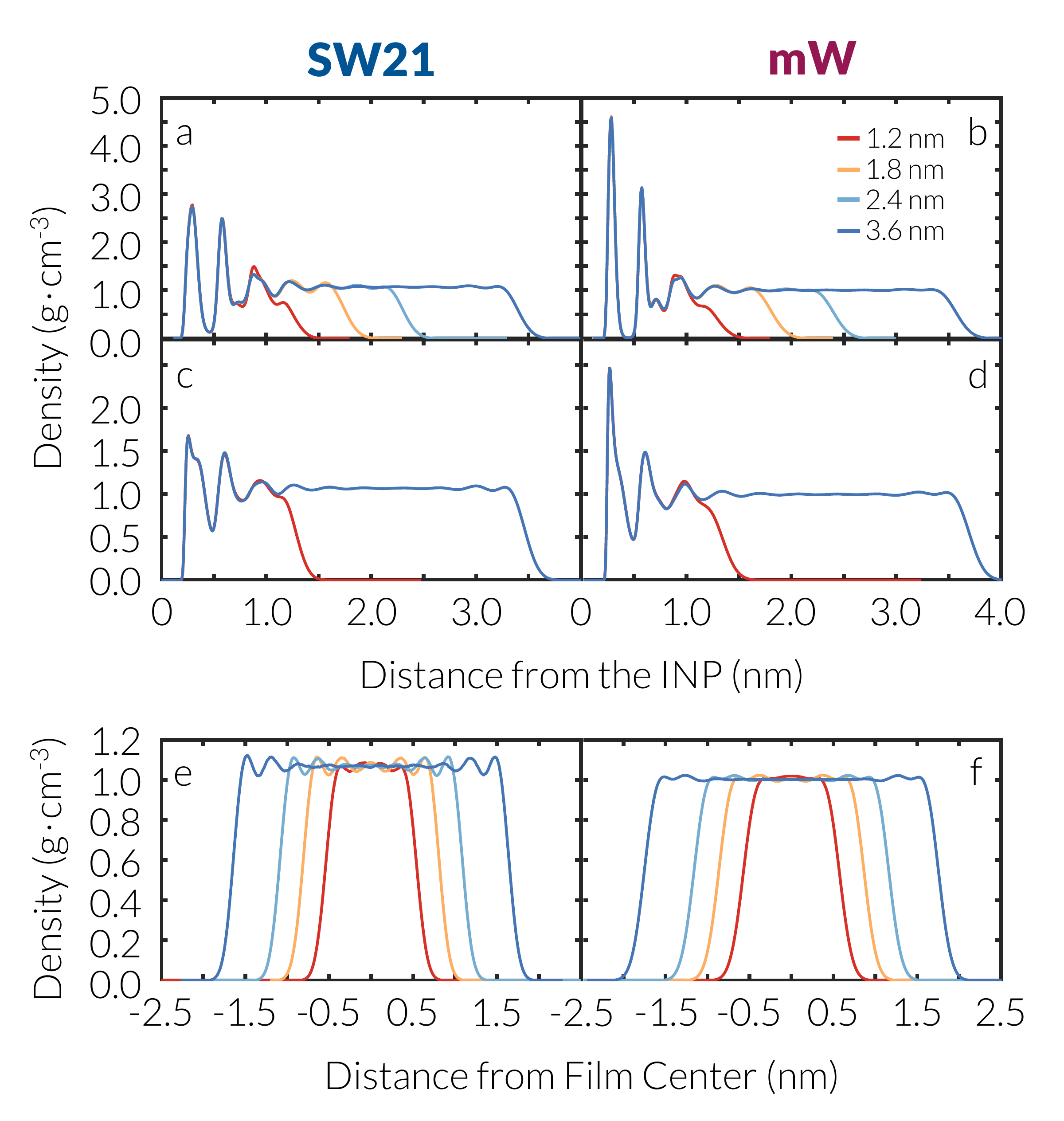} 
\caption{Density profiles for (a-b) graphene-supported films, (c-d) supported films in the vicinity of an LJ 9-3 structureless wall, and (e-f) freestanding films. (a,c,e) correspond to films of SW21 while (b,d,f) are for mW films. For graphene-supported and freestanding films, density profiles are computed for four film thicknesses, 1.2, 1.8, 2.4~and~3.6~nm. For supported films next to an LJ 9-3 structureless wall, only two films of thicknesses 1.2 and 3.6~nm are considered.}
\label{SI-fig:density}
\end{figure*} 

\begin{figure*}[h]
\centering
\includegraphics[width=0.424\textwidth]{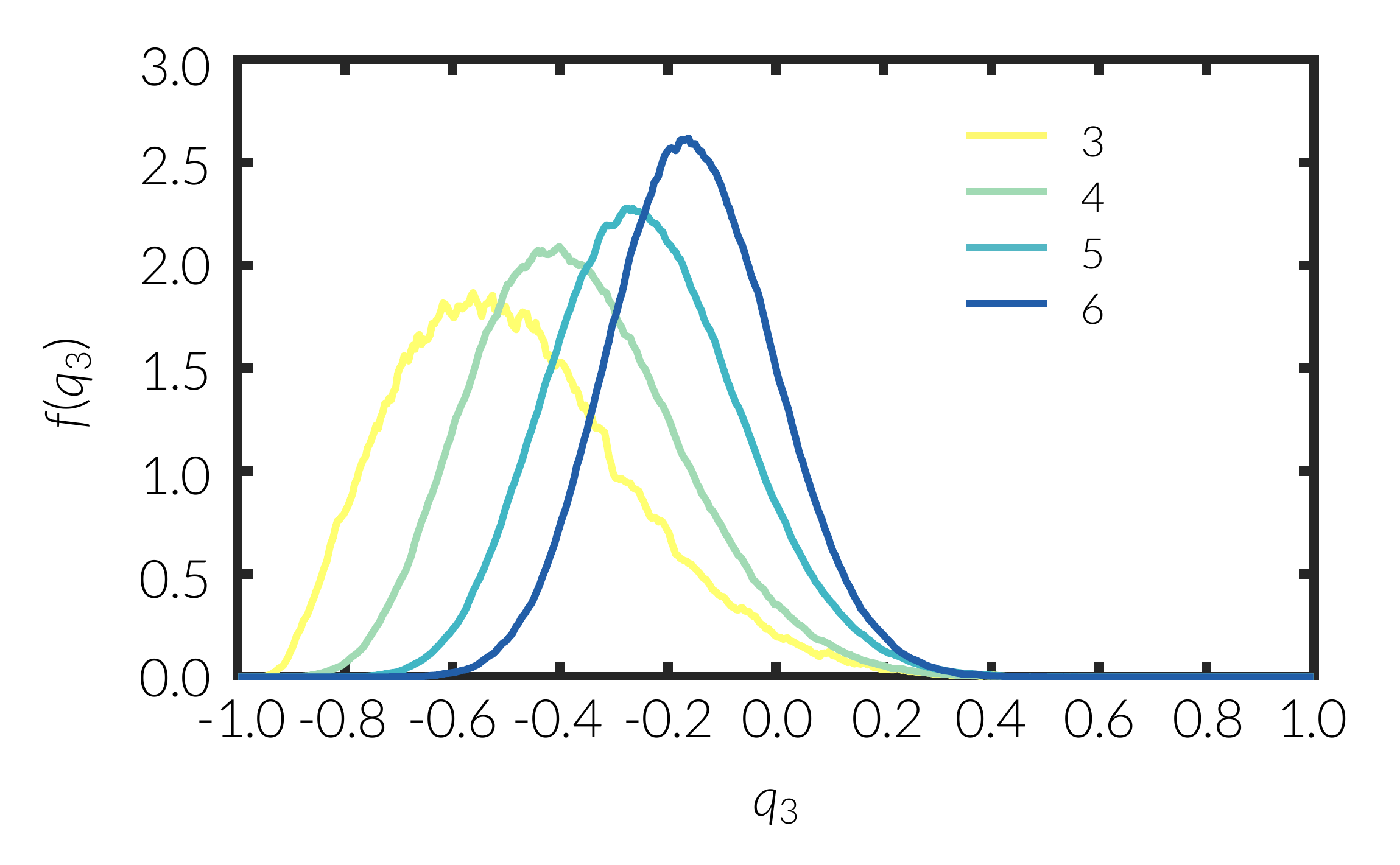}
\caption{$\textit{q}_\text{3}$ distributions computed for molecules with a fixed number of nearest neighbors within a 2.4-nm thick SW21 film at \text{T}=170~K.}
\label{SI-fig:q3-ngb}
\end{figure*}

\begin{figure*}[h]
\centering
\includegraphics[width=0.4755\textwidth]{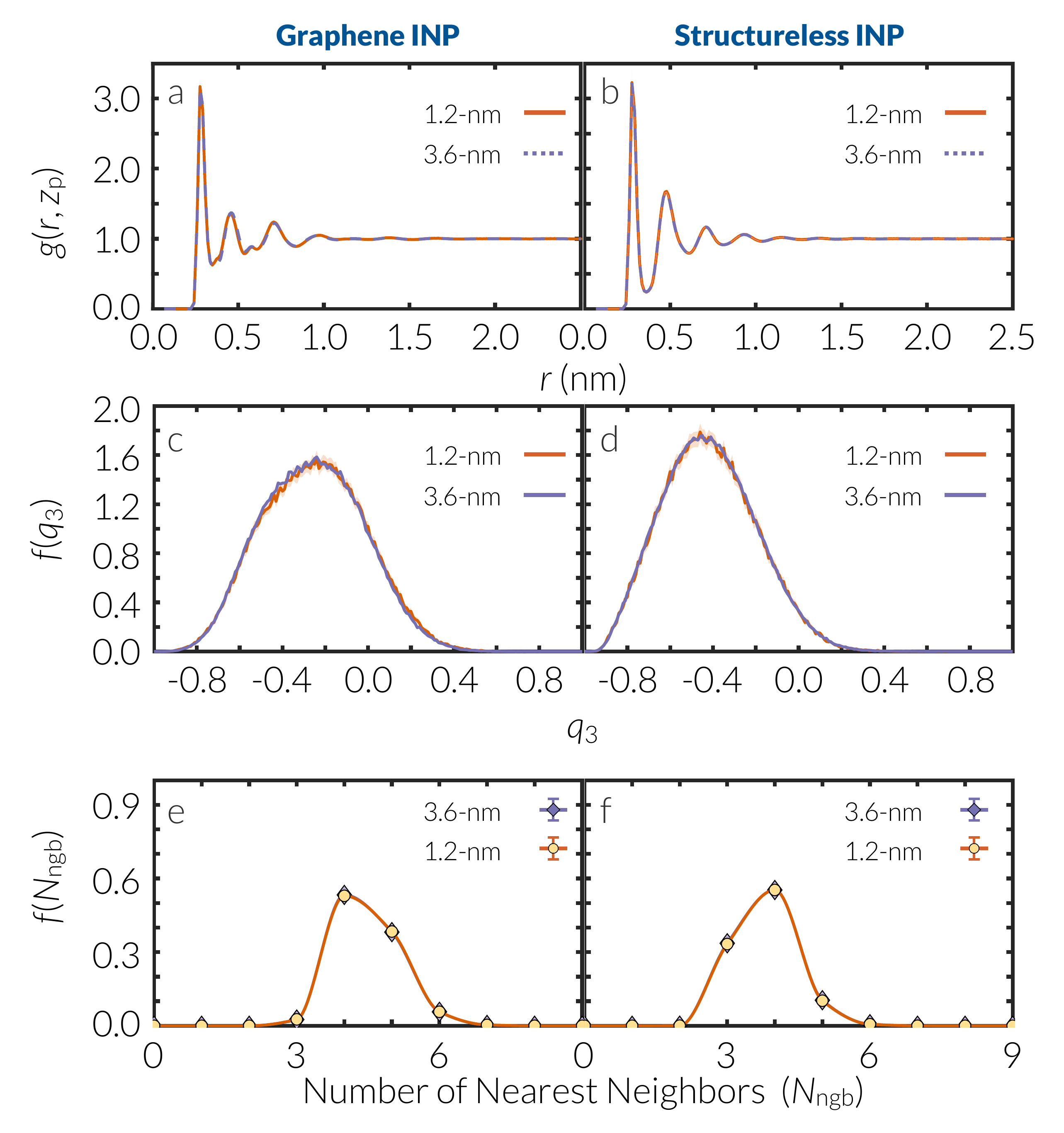}
\caption{Structural properties of the INP-adjacent interfacial region of  supported SW21 films in the vicinity of (a,c,e) graphene and (b,d,f) LJ 9-3 structureless wall. (a-b) Planar RDF's; (c-d) $\textit{q}_\text{3}$ distributions, and (e-f) nearest neighbor count distributions. Structural features close to the wall do not vary between the ultrathin and thicker films, suggesting that the free interface does not modulate the structure of the INP-adjacent interface. {\color{black}Error bars in (a,b) and (e,f) are thinner than the curves and smaller  than the symbols, respectively. The areas under the curves in (c,d) are normalized to unity, and the shades correspond to error bars.}}
\label{SI-fig:INP-adjacent}
\end{figure*} 

\begin{figure*}[h]
\centering
\includegraphics[width=0.9595\textwidth]{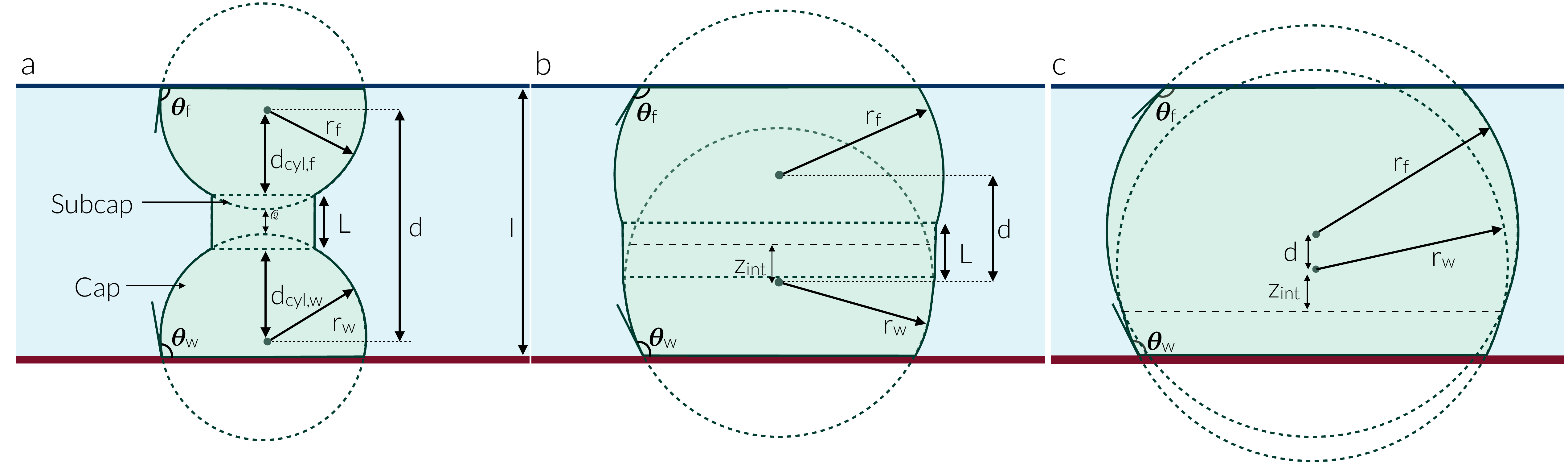} 
\caption{Different scenarios for the relative arrangements of the two spherical caps within an hourglass-shaped nucleus. The two caps do not intersect in (a) while they intersect in (b) and (c). (b) and (c) differ in terms of the relative position of the intersecting plane, which is in between the centers of the two spheres in (b), but not in (c).  A liquid-exposed cylindrical bridge is only possible in (a) and (b).  }
\label{SI-fig:schematic-hourglass} 
\end{figure*}

\begin{figure*}[h]
\centering
\includegraphics[width=0.4005\textwidth]{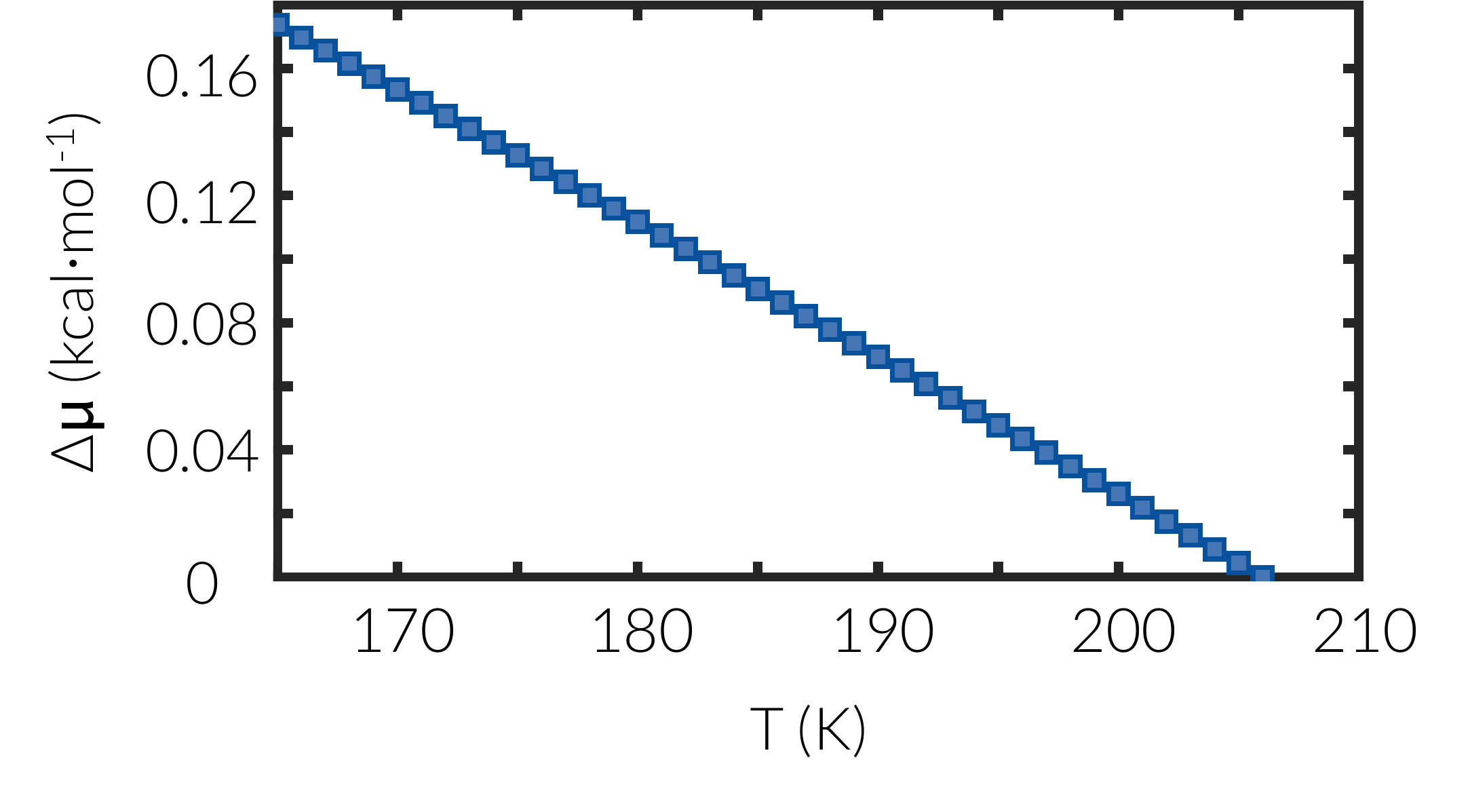}
\caption{$\Delta\mu$ as a function of $\textit{T}$ in the SW21 system.}
\label{SI-fig:dmuvsT}
\end{figure*}

\begin{table*}
\centering
\caption{\color{black}The confidence intervals for predictions of the CNT-based theory obtained via sensitivity analysis. Note that nucleation is not feasible for the unperturbed $\theta_\textit{f}$ in the vicinity of a structureless INP, so no estimates for $\Delta{\textit{G}}_{\text{diff,orig}}$ {\color{black} and $N^*_{\text{orig}}$} are  available. \label{tab:theory-errorbars}}
\begin{tabular}{cccc}
\hline\hline
\color{black}System & \color{black}Property & \color{black}Best Estimate & \color{black}Confidence Interval \\
\hline
\color{black}SW21/Graphene & \color{black}$\Delta{\textit{G}}_{\text{diff,orig}}$ & \color{black}5.98 & \color{black}[2.39, 9.37]\\
\color{black}SW21/Graphene & \color{black}$\textit{N}^*_{\text{orig}}$ & \color{black} 216 & \color{black}[137, 329]\\
\color{black}SW21/Graphene & \color{black}$\theta_{\textit{f},\text{corr}}$ & \color{black}118.6$^\circ$ & \color{black}[98.0$^{\circ}$, 132.7$^{\circ}$] \\
\color{black}SW21/Graphene & \color{black}$\textit{N}^*_{\text{corr}}$ & \color{black}150 & \color{black}[75, 296] \\
\hline
\color{black}SW21/LJ 9-3 & \color{black}$\Delta{\textit{G}}_{\text{diff,orig}}$ & \color{black}N/A & \color{black}N/A \\
\color{black}SW21/LJ 9-3 & \color{black}$\textit{N}^*_{\text{orig}}$ & \color{black} N/A & \color{black} N/A\\
\color{black}SW21/LJ 9-3 & \color{black}$\theta_{\textit{f},\text{corr}}$ & \color{black}94.6$^{\circ}$ & \color{black}[75.7$^\circ$, 109.2$^\circ$]\\
\color{black}SW21/LJ 9-3 & \color{black}$\textit{N}^*_{\text{corr}}$ & \color{black}124 & \color{black}[89, 282]\\
\hline\hline
\end{tabular}
\end{table*}



\end{document}